\documentclass[12pt]{article}

\newtheorem{one}{Proposition}[section]{\bf}{\em}
\newtheorem{two}[one]{Proposition}{\bf}{\em}
\newtheorem{three}{Proposition}[section]{\bf}{\em}
\newtheorem{four}[three]{Proposition}{\bf}{\em}
\newtheorem{five}[three]{Corollary}{\bf}{\em}
\newtheorem{six}[three]{Proposition}{\bf}{\em}
\newtheorem{seven}[three]{Corollary}{\bf}{\em}

\usepackage{amssymb}

\begin{document}
\title{Conservation laws for
a class of nonlinear equations with variable coefficients
on discrete and noncommutative spaces}
\author{M. Klimek \\
Institute of Mathematics and Computer Science, \\ Technical University
of Cz\c{e}stochowa, \\ ul.D\c{a}browskiego 73, 42-200 Cz\c{e}stochowa, Poland \\
klimek@matinf.pcz.czest.pl}
\maketitle

\begin{abstract}
The conservation laws for a class of nonlinear equations
with variable coefficients on discrete and noncommutative
spaces are derived. For discrete models the conserved charges
are constructed explicitly. The applications of the general method include
equations on quantum plane, supersymmetric equations for chiral and antichiral
supermultiplets as well as auxiliary equations of integrable models - principal chiral model
and various cases of nonlinear Toda lattice equations.
\end{abstract}
\section{Introduction}
Our aim is to present the procedure of derivation of the conservation laws and consequently
the conserved charges for a certain class of nonlinear equations with variable coefficients.
In the classical field theory the conserved currents and charges follow from the Noether
theorem and are connected with the symmetry of the action. In the case of linear equations of motion
with constant coefficients
the construction of conserved currents by Takahashi and Umezawa method \cite{a0} can be applied.\\
In the previous papers \cite{b,c,c1,c2} we have extended this procedure to the linear equations on discrete
and noncommutative spaces including quantum Minkowski \cite{r,s,o} and braided linear spaces \cite{n,m,l,Ma}.
It appears however that we can consider in a similar way
a wide class of equations with variable coefficients built within framework of any differential calculus with the Leibnitz rule
for partial derivatives  deformed via the transformation operator which is multiplicative
 and invertible. The range of admissible spaces
includes the classical space-time with continuous coordinats, the discrete space,
the mixed space with discrete and continuous coordinates, superspace including space-time
and spinor coordinates, quantum Minkowski and braided linear spaces (with q-Minkowski as a special case).\\
The equations we shall investigate have the variable coefficients which
fulfill the corresponding restriction (\ref{cond}) including the conjugated derivatives.
The possibility of derivation the conservation laws for some equations
with variable coefficients on
noncommutative spaces was indicated earlier \cite{c1,c2} but the proposed conditions for coefficients
were too strong to construct usefull examples. In constrast the new condition (as we show
explicitly in the applications) appears to be identical
with the nonlinear equations of some of the integrable models for which we
derive the conserved currents via
the auxiliary linear equations with variable coefficients. \\
Resuming the proposed method can be applied to nonlinear models in two ways namely we can consider
the equation with nonlinear term free from the derivatives (in form of the potential)
or alternatively  we investigate the auxiliary
linear equations with variable coefficients for nonlinear integrable models. \\
The paper is organized as follows: the section 2 contains the description of the
investigated models and the derivation of the conservation laws. In section 3 we extend the procedure to discrete
and mixed discrete and continuous models including both types of derivatives
initial $\partial $ and its conjugation $ \partial^{\dagger}$.
 For this class of equations we add the construction of conserved charges
 following the classical method of integrating the time-component  of the
 conserved current over the subspace excluding the time-coordinate. This result is due to the fact
 that for the discrete and classical differential calculi the definite integral
 over mentioned subspace is known to commute with time-derivative (\ref{sub}). \\
 The section 4 includes applications of the derived procedure. We start
 with a simple example of an equation of the second order with variable coefficients
 on quantum plane, then we consider the conserved currents and charges
 for a pair of chiral and antichiral supermultiplts on D=4 N=1 superspace connected
 via nonlinear equation of motion. This section is closed with a review
 of conservation laws and conserved charges for a range of integrable models.
 We apply our method to the principal chiral model equations and their auxiliary
 linear equations. In this example only the classical commutative derivatives
 are involved; nevertheless it illustrates well the connection between the imposed condition (\ref{cond})
 and the initial nonlinear (in this case the principal chiral model) equation. The next model is the nonlinear
 Toda lattice equation for which we use our results for mixed discrete and continuous spaces
 in order to write explicitly the conserved currents and charges.
  \section{Nonlinear equations with variable coefficients and their conservation
law}
In the previous papers we discussed the conservation laws for linear equations
with constant coefficients for discrete differential calculus \cite{b,c} as well as
for a wider category of equations acting on  noncommutative spaces namely on quantum Minkowski spaces \cite{c1,d}
and the braided linear ones \cite{c2,c3,c4}. \\
Now we would like to present our results for an extended class of equations
with variable coefficients of the form:
\begin{eqnarray}
  & \Lambda (\partial) \Phi=0 & \label{eq}      \\
  & \Lambda (\partial)=\Lambda_{0}+ \sum_{l=1}^{N} \Lambda_{\mu_{1}... \mu_{l}} \label{lambda}
  \partial^{\mu_{1}}...\partial^{\mu_{l}} & \label{operator}
\end{eqnarray}
where coefficients (may be matrices) fulfill the condition:
\begin{equation}
\partial^{\dagger \;\; \mu_{1}} \Lambda_{ \mu_{1}...\mu_{l}}=0
\label{cond}
\end{equation}
for $ l=1,...,N $ and  with the conjugated derivative described below (\ref{conj}). \\
We include into considerations the nonlinear equations provided the only
nonlinear term does not depend on derivatives that means:
\begin{equation}
              \Lambda_{0}=\Lambda_{0}(\Phi)
              \label{nonlinear}
\end{equation}
As will be shown in the sequel this class of equations contains the equations of motion for
supersymmetric models in superfield formulation as well as auxiliary linear equations yielded
by nonlinear integrable models as for example Toda lattice equations or principal
chiral model equations (the last being an example built using classical commutative derivatives). \\
Our construction presented below in Propositions  2.1 and 2.2 holds for any differential calculus
with the following deformation of the Leibnitz rule:
\begin{equation}
\partial^{i}(fg)=(\partial^{i}f)g+(\zeta^{i}_{j}f)\partial^{j}g
\label{leib}
\end{equation}
where $ \zeta $ is the invertible transformation operator. \\
Starting with the classical commutative differential calculus (for which $ \zeta $
is the identity operator) we can include the discrete differential calculus with derivatives
defined by one of the formulas:
\begin{eqnarray}
& \partial^{i} f_{k}=(\zeta^{i}-1)f_{k}=f_{k+1}-f_{k} &  \label{discrete}\\
& \partial^{i} f_{k}=(x_{k+1}-x_{k})^{-1}(\zeta^{i}-1)f_{k}=
(x_{k+1}-x_{k})^{-1}(f_{k+1}-f_{k}) & \label{discrete1}
\end{eqnarray}
where $ k $ denotes the position on the lattice in the direction $ i $. \\
The first one is widely used in the discrete models and measures the difference
between the values of the function between two points of the lattice in the given direction $ i $ while the second one (introduced \cite{b,c})
is a quotient of the change of value of the function over the distance between the
two neighbouring points of the lattice taken in the direction $ i $.
Both types of the derivatives obey (\ref{leib}) with the transformation
operator being the shift operator along the lattice in the direction $ i $:
\begin{equation}
\zeta^{i} f_{k}=f_{k+1}
\end{equation}
where $ k $ in the above formula denotes the value of the $i$-th coordinate
of the point on the lattice.\\
 It is clear that we can also include
mixed models depending both on classical commutative derivatives with respect to continuous
coordinates and discrete derivatives (\ref{discrete}) or (\ref{discrete1}) with respect to lattice dimensions. \\
The Leibnitz rule of the form (\ref{leib}) is also characteristic for noncommutative spaces.
We have checked \cite{c1,c2} that for quantum Minkowski spaces and for braided linear spaces (including the q-Minkowski space)
it is given by the formula (\ref{leib}) with transformation operators determined by their multiplicity property:
\begin{equation}
\zeta^{i}_{j}(fg)=(\zeta^{i}_{k}f)(\zeta^{k}_{j}g)
\label{mult}
\end{equation}
and the action on monomials of the first order:
\begin{eqnarray}
& \zeta^{i}_{j}x^{k}=R^{ik}_{aj}x^{a}-(RZ)^{ik}_{j} & \label{for1} \\
& \zeta^{i}_{j}x_{k}=R^{li}_{kj}x_{l}&   \label{for2}
\end{eqnarray}
The first of the above formulas is valid on the quantum Minkowski  and the second
on the braided linear space with $ R $ being the matrices fulfilling the QYBE \cite{o,n}. For quantum Minkowski spaces introduced by Podle\'{s}
and Woronowicz the $ R $ matrix is selfinvertible $ R^{2}=1 $ and for braided linear spaces developed
by Majid it is bi-invertible:
\begin{eqnarray}
& (R^{-1})^{ij}_{kl} R^{kl}_{ab}=R^{ij}_{kl} (R^{-1})^{kl}_{ab}=\delta^{i}_{a}\delta^{j}_{b} & \\
& \tilde{R}^{ib}_{al} R^{ak}_{jb}= R^{ib}_{al} \tilde{R}^{ak}_{jb}=\delta^{i}_{j}\delta^{k}_{l} &
\end{eqnarray}
Let us notice that the important category  of models on noncommutative spaces
are supersymmetric models in the superspace formulation. In this framework the noncommutative
space is divided into the classical commutative Minkowski space  coordinates $ x $
(with corresponding indices and metric) and the spinor coordinates $ \theta $
(we shall not specify below the type of spinors - Majorana or Weyl). For all
equations within this class the part connected with space-time is described
using the classical commutative differential calculus while for spinor
coordinates we have anticommuting derivatives. The first part of the components
of the transformation operator
looks as follows:
\begin{equation}
\zeta^{\mu}_{\nu}=\delta^{\mu}_{\nu} \hspace{1cm}
\zeta^{\alpha}_{\nu}=\zeta^{\nu}_{\alpha}=0
\end{equation}
where for the supersymmetric models we have denoted the space-time indices as $ \mu $, $ \nu$
while we use $ \alpha $ and $ \beta $ as spinor indices (Weyl or Majorana). The only
nontrivial part of the transformation operator for the supersymmetric models
are the components which act between the spinor indices namely for the monomials
of the first order they are of the form:
\begin{equation}
\zeta^{\alpha}_{\beta} x_{\mu}=\delta^{\alpha}_{\beta} x_{\mu} \hspace{2cm}
\zeta^{\alpha}_{\beta} \theta_{\gamma}=-
\delta^{\alpha}_{\beta}\theta_{\gamma}
\end{equation}
We can extend the supersymmetric transformation operator to an arbitrary function
by its multiplicity property (\ref{mult}). \\
For all the considered differential calculi  (discrete, supersymmetric,
quantum Minkowski and braided) the transformation operators have
one common feature namely they are invertible; for discrete models with derivatives (\ref{discrete})
or (\ref{discrete1}) the inverse operators are simply backshift operators in the given direction
while for noncommutative models the inverses are given by their multiplicity property:
\begin{equation}
\zeta^{-\;\;\; i}_{j}(fg)=(\zeta^{- \;\;\; k}_{j}f)(\zeta^{-\;\;\; i}_{k}g)
\label{mult1}
\end{equation}
together with their action on monomials of the first order:
\begin{eqnarray}
& \zeta^{-\;\;\; i}_{j}x^{k}=R^{ki}_{ja}x^{a}+Z^{ki}_{j} & \label{for1'}\\
& \zeta^{-\;\;\; i}_{j}x_{k}=\tilde{R}^{li}_{kj}x_{l}&   \label{for2'}
\end{eqnarray}
where $ \tilde{R} $ is the second inverse of the $ R $ matrix characteristic for braided
linear spaces \cite{n}. \\
The inverse operator for supersymmetric models  is given by:
\begin{equation}
\zeta^{-\;\; \mu}_{\nu}=\delta^{\mu}_{\nu} \hspace{1cm}
\zeta^{-\;\; \alpha}_{\nu}=\zeta^{-\;\; \nu}_{\alpha}=0
\end{equation}
with the spinor - spinor part defined by the multiplicity (\ref{mult1}) and its action on monomialas
of the first order:
\begin{equation}
\zeta^{- \;\; \alpha}_{\beta} x_{\mu}=\delta^{\alpha}_{\beta} x_{\mu} \hspace{2cm}
\zeta^{-\;\; \alpha}_{\beta} \theta_{\gamma}=-
\delta^{\alpha}_{\beta}\theta_{\gamma}
\end{equation}
All the described above inverse operators fulfill the condition:
\begin{equation}
\zeta^{k}_{j}\zeta^{- \;\;\;i}_{k}=\zeta^{-\;\;\; k}_{j}\zeta^{i}_{k}=\delta^{i}_{j}
\label{inverse}
\end{equation}
The properties of the transformation operators $ \zeta $ and $ \zeta^{-} $
imply the following modification of the Leibnitz rule:
\begin{equation}
\partial^{k}[(\zeta^{-\;\;\; i}_{k} f)g]=(-\partial^{\dagger \;\;\; i}f)g
+f(\partial^{i}g)=f \left (- \stackrel{\leftarrow}{\partial}^{\dagger \;\;\; i}
+\partial^{i} \right ) g
\label{leibmod}
\end{equation}
where the conjugated derivative $ \partial^{\dagger} $ is defined as follows:
\begin{equation}
\partial^{\dagger \;\;\; i}:=-\partial^{k}\zeta^{-\;\;\; i}_{k}
\label{conj}
\end{equation}
and we take for the given type of the derivative the connected
inverse transformation operator.
We see that after modification we deal with the Leibnitz rule where the
right-hand side  is analogous to the classical differential calculus:
$$ \partial^{i} (fg)=f \left (-\stackrel{\leftarrow}{\partial}^{\dagger \;\;\; i}
+\partial^{i} \right ) g $$
for which the conjugated derivative is given by: $ \partial^{\dagger \;\; i}=-\partial^{i} $. \\
The conjugated derivatives form the conjugated equation which for the class
described by conditions (\ref{eq} - \ref{cond}) looks as follows:
\begin{equation}
\Lambda(\stackrel{\leftarrow}{\partial}^{\dagger})= \Lambda_{0} (\Phi)
+ \sum_{l=1}^{N}
  \stackrel{\leftarrow}{\partial}^{\dagger \;\; \mu_{1}}...
  \stackrel{\leftarrow}{\partial}^{\dagger \;\; \mu_{l}}
   \left(\zeta^{- \;\; \alpha_{l}}_{\mu_{l}}...\zeta^{- \;\; \alpha_{1}}_{\mu_{1}}\Lambda_{\alpha_{1}... \alpha_{l}}\right)
   \label{eqconj}
\end{equation}
The following propositions hold for equation (\ref{eq} - \ref{cond}) and are the extension of results
derived in \cite{c,c1,c2}.
\begin{one}
The unique solution of the operator equation:
\begin{equation}
\sum_{\mu} (-\stackrel{\leftarrow}{\partial}^{\dagger \;\; \mu} +\partial^{\mu} ) \circ
\Gamma_{\mu} (\partial, \stackrel{\leftarrow}{\partial}^{\dagger})=
 \Lambda(\partial)- \Lambda(\stackrel{\leftarrow}{\partial}^{\dagger})
 \label{cond1}
\end{equation}
in the class of polynomials of derivatives
$ \stackrel{\leftarrow}{\partial}^{\dagger} $ and $ \partial $
for the equation operator $ \Lambda $ fulfilling (\ref{cond})
is of the form:
\begin{equation}
\Gamma_{\mu} (\partial, \stackrel{\leftarrow}{\partial}^{\dagger})=
(\zeta^{-\;\; \alpha}_{\mu}\Lambda_{\alpha})+
\label{gamma}
\end{equation}
$$
+\sum_{l=1}^{N-1} \sum_{k=0}^{l}
\stackrel{\leftarrow}{\partial}^{\dagger \;\; \mu_{1}}...
\stackrel{\leftarrow}{\partial}^{\dagger \;\; \mu_{k}}
\left (\zeta^{-\;\; \alpha}_{\mu}\zeta^{-\;\; \alpha_{k}}_{\mu_{k}}...\zeta^{-\;\; \alpha_{1}}_{\mu_{1}}
\Lambda_{\alpha_{1}...\alpha_{k} \alpha \mu_{k+1}...\mu_{l}}\right)
\partial^{\mu_{k+1}}... \partial^{\mu_{l}}
$$
\end{one}
Proof:\\
Let us explain that
the "$ \circ $" operation describes the way the noncommuting derivatives work on
monomials of derivatives $ \stackrel{\leftarrow}{\partial}^{\dagger} $
and $ \partial$, namely:
\begin{eqnarray}
 & (-\stackrel{\leftarrow}{\partial}^{\dagger \;\; \mu} +\partial^{\mu} ) \circ
\overline{[\nu_{1},...,\nu_{l}]}a(\vec{x})[\rho_{1},...,\rho_{k}]:= & \nonumber \\
 & - \overline{[\nu_{1},...,\nu_{l},\mu]}a(\vec{x})[\rho_{1},...,\rho_{k}]
+\overline{[\nu_{1},...,\nu_{l}]}\partial^{\mu} a(\vec{x})[\rho_{1},...,\rho_{k}] &
\end{eqnarray}
where the following notation for monomials was used:
\begin{eqnarray}
& [\rho_{1},...,\rho_{k}]:=\partial^{\rho_{1}}... \partial^{\rho_{k}} &   \label{monomial1}\\
& \overline{[\nu_{1},...,\nu_{l}]}:=
\stackrel{\leftarrow}{\partial}^{\dagger \;\; \nu_{1}}...
\stackrel{\leftarrow}{\partial}^{\dagger \;\; \nu_{l}} & \label{monomial2}
\end{eqnarray}
The further calculations base on the assumption that we consider the general
operator of the order $ N-1 $ with respect to the derivatives
and on the associativity of the algebra of derivatives. The explicit
 solution of the condition (\ref{cond1}) is
enclosed in Appendix A. 
\\
We modify the $ \Gamma $ operator due to the deformation of the Leibnitz rule (\ref{leib}, \ref{leibmod})
and obtain the operator $ \hat{\Gamma} $:
\begin{equation}
\hat{\Gamma}_{\mu} (\partial, \stackrel{\leftarrow}{\partial}^{\dagger})=
\stackrel{\leftarrow}{\zeta}^{- \;\;\; j}_{\mu}
(\zeta^{-\;\; \alpha}_{j}\Lambda_{\alpha})+
\label{hgamma}
\end{equation}
$$
 +\sum_{l=1}^{N-1} \sum_{k=0}^{l}
\stackrel{\leftarrow}{\partial}^{\dagger \;\; \mu_{1}}...
\stackrel{\leftarrow}{\partial}^{\dagger \;\; \mu_{k}}
\stackrel{\leftarrow}{\zeta}^{-\;\;\; j}_{\mu}
\left (\zeta^{- \;\; \alpha}_{j} \zeta^{- \;\; \alpha_{k}}_{\mu_{k}}...\zeta^{- \;\; \alpha_{1}}_{\mu_{1}}\Lambda_{\alpha_{1}...\alpha_{k} \alpha \mu_{k+1}...\mu_{l}}\right )
\partial^{\mu_{k+1}}... \partial^{\mu_{l}}
$$
which for a pair of arbitrary functions $ f $ and $ g $ is connected with the $ \Gamma $
operator by the equality:
\begin{equation}
\sum_{\mu}\partial^{\mu} f \hat{\Gamma}_{\mu}(\partial, \stackrel{\leftarrow}{\partial}^{\dagger}) g=
\sum_{\mu} f\left (-\stackrel{\leftarrow}{\partial}^{\dagger \;\;\; \mu}+\partial^{\mu}
\right )\circ \Gamma_{\mu}(\partial, \stackrel{\leftarrow}{\partial}^{\dagger}) g
\label{hgamma1}
\end{equation}
The immediate consequence of the Proposition  2.1 is therefore
 the construction of conserved currents yielded
by the following statement.
\begin{two}
Let us assume that function $ \Phi $ is an arbitrary solution of equation (\ref{eq}, \ref{operator})
with coefficients fulfilling (\ref{cond}), that means:
\begin{equation}
\Lambda (\partial) \Phi =0
\end{equation}
and function $ \Phi' $ solves the conjugated equation with the operator of the form (\ref{eqconj}):
\begin{equation}
\Phi' \Lambda (\stackrel{\leftarrow}{\partial}^{\dagger}) =0
\label{eq1}
\end{equation}
Then
\begin{equation}
J_{\mu}=\Phi' \hat{\Gamma}_{\mu} (\partial, \stackrel{\leftarrow}{\partial}^{\dagger})\Phi
\label{current}
\end{equation}
where the operator $ \hat{\Gamma}_{\mu} $  is defined by (\ref{hgamma}), is a current which
obeys the conservation law:
\begin{equation}
\sum_{\mu} \partial^{\mu} J_{\mu}=0
\label{cons}
\end{equation}
\end{two}
Proof: \\
This is a corollary from the (\ref{hgamma1}) property of the $ \hat{\Gamma} $ operator, from the operator equation (\ref{cond1})
for $ \Gamma $ and finally from the fact that functions $ \Phi $ and $ \Phi' $
fulfill the respective equations (\ref{eq}, \ref{eq1}). 
\\
Let us notice that the auxiliary conjugated equation for nonlinear model is now the linear one
with respect to the solution $ \Phi'$:
\begin{equation}
\Phi'  \left ( \Lambda_{0} (\Phi)
+ \sum_{l=1}^{N}
  \stackrel{\leftarrow}{\partial}^{\dagger \;\; \mu_{1}}...
  \stackrel{\leftarrow}{\partial}^{\dagger \;\; \mu_{l}}
   \zeta^{- \;\; \alpha_{l}}_{\mu_{l}}...\zeta^{- \;\; \alpha_{1}}_{\mu_{1}}\Lambda_{\alpha_{1}... \alpha_{l}}\right )=0
\end{equation}
The solution $ \Phi' $ depends on the solution of the initial equation $ \Phi $
which defines the potential for the conjugated equation $ \Lambda_{0} (\Phi ) $. \\
Now the interesting question arises whether the presented construction which works
on a wide class of spaces can be extended to equations with operators including both $ \partial $ and $ \partial^{\dagger} $
derivatives. As we know from discrete models (see for example \cite{mh1,mh2}) the equations depend explicitly
on the forth- and backshifts along the lattice. Also the equations of motion built
within the framework of generalized difference derivatives (\ref{discrete1}) (which was discussed in \cite{a,c} )
include both initial and conjugate derivatives as a consequence of the minimal action
principle. The method of derivation of the conserved currents for the discrete and mixed models
shall be described in the next section. It is based on the fact that acting on functions
of commutative coordinates the operators (\ref{discrete}) and (\ref{discrete1}) yield two symmetric formulas for the Leibnitz rules
(both for initial and conjugated derivative).
\section{Conserved currents and charges for discrete and mixed models depending on the $ \partial$ and
$ \partial^{\dagger}$ derivatives}
Our aim is now to extend the construction described in Propositions  2.1 and 2.2 to the equations depending on the
initial and conjugated derivatives of the form:
\begin{eqnarray}
              & [\Lambda(\partial) + \tilde{\Lambda}(\partial^{\dagger})] \Phi=0 & \label{eqbis}\\
              & \Lambda(\partial)=\Lambda_{0}+\sum_{l=1}^{N} \Lambda_{\mu_{1}...\mu_{l}}
              \partial^{\mu_{1}}...\partial^{\mu_{l}} & \\
              & \tilde{\Lambda}(\partial^{\dagger})=\tilde{\Lambda}_{0}+\sum_{l=1}^{\tilde{N}} \tilde{\Lambda}_{\mu_{1}...\mu_{l}}
              \partial^{\dagger \;\; \mu_{1}}...\partial^{\dagger \;\; \mu_{l}} & \label{operator1}
\end{eqnarray}
Following the previous considerations we restrict the class of equations to variable coefficients
fulfilling the extended version of the condition (\ref{cond}):
\begin{equation}
              \partial^{\dagger \;\; \mu_{1}}\Lambda_{\mu_{1}...\mu_{l}}=0
              \hspace{1cm}
              \partial^{\mu_{1}}\tilde{\Lambda}_{ \mu_{1}...\mu_{k}}=0
              \label{condconj}
\end{equation}
for all $ l=1,...,N$ and $ k=1,...,\tilde{N} $. \\
We include the nonlinear equations provided the nonlinear terms do not depend on the derivatives that means
they have a form of potential:
\begin{equation}
              \Lambda_{0}=\Lambda_{0}(\Phi) \hspace{2cm}   \tilde{\Lambda}_{0}=\tilde{\Lambda}_{0}(\Phi)
              \label{nonlinear1}
\end{equation}
As we shall see in the subsequent sections the class of equations fufilling (\ref{condconj})
contains the widely studied in different context linear auxiliary equations connected with
nonlinear integrable models such as principal chiral model and equations
of nonlinear Toda lattice models.\\
The important factor in our calculations is the fact that for discrete (or mixed -
classical and discrete) differential calculus we can write the modified Leibnitz rule
symmetricaly in the following form:
\begin{eqnarray}
              & \partial^{\alpha}(\zeta^{- \;\; \beta}_{\alpha} f)g=
              (-\partial^{\dagger \;\; \beta}f)g+f\partial^{\beta}g=
              f(-{\stackrel{\leftarrow}{\partial}}^{\dagger \;\; \beta}+\partial^{\beta})g & \label{leib1} \\
              & \partial^{\dagger \;\; \alpha}(\zeta^{ \beta}_{\alpha} f)g=
              (-\partial^{\beta}f)g+f\partial^{\dagger \;\; \beta}g=
              f(-{\stackrel{\leftarrow}{\partial}}^{\beta}+\partial^{\dagger \;\; \beta})g & \label{leib2}
\end{eqnarray}
The second of the above equations indicates that the statement analogous to the Proposition  2.1
holds for the operator $ \tilde{\Lambda}(\partial^{\dagger})$; namely we shall prove that the operator equation
given below has the unique solution by virtue of Proposition 3.1:
\begin{equation}
              \sum_{\mu} (-\stackrel{\leftarrow}{\partial}^{\mu} +\partial^{\dagger \;\; \mu} ) \circ
\tilde{\Gamma}_{\mu} (\partial^{\dagger}, \stackrel{\leftarrow}{\partial})=
 \tilde{\Lambda}(\partial^{\dagger})- \tilde{\Lambda}(\stackrel{\leftarrow}{\partial})
 \label{cond1bis}
\end{equation}
where the conjugated operator $ \tilde{\Lambda}(\stackrel{\leftarrow}{\partial}) $
looks as follows:
\begin{equation}
\tilde{\Lambda}(\stackrel{\leftarrow}{\partial})=\tilde{\Lambda}_{0}(\Phi)+
\sum_{l=1}^{\tilde{N}}
\stackrel{\leftarrow}{\partial}^{\mu_{1}}...\stackrel{\leftarrow}{\partial}^{\mu_{l}}
\left (\zeta^{\alpha_{l}}_{\mu_{l}}...\zeta^{\alpha_{1}}_{\mu_{1}}
\Lambda_{\alpha_{1}...\alpha_{l}} \right )
\label{eqconj1}
\end{equation}
\begin{three}
The unique solution of (\ref{cond1bis}) in the class of polynomials of derivatives
$ \stackrel{\leftarrow}{\partial} $ and $ \partial^{\dagger} $
for the equation operator $ \tilde{\Lambda} $ (\ref{operator1})  with  coefficients
fulfilling (\ref{condconj})
is of the form:
\begin{equation}
\tilde{\Gamma}_{\mu} (\partial^{\dagger}, \stackrel{\leftarrow}{\partial})=
\label{tgamma}
\end{equation}
$$
(\zeta^{\alpha}_{\mu}\tilde{\Lambda}_{\alpha}) +\sum_{l=1}^{\tilde{N}-1} \sum_{k=0}^{l}
\stackrel{\leftarrow}{\partial}^{ \mu_{1}}...
\stackrel{\leftarrow}{\partial}^{ \mu_{k}}
\left (\zeta^{\alpha}_{\mu}  \zeta^{ \alpha_{k}}_{\mu_{k}}... \zeta^{ \alpha_{1}}_{\mu_{1}}
\tilde{\Lambda}_{\mu_{1}...\mu_{k} \alpha \mu_{k+1}...\mu_{l}} \right )
\partial^{\dagger \;\; \mu_{k+1}}... \partial^{\dagger\;\; \mu_{l}}
$$
\end{three}
Proof:\\
The formula (\ref{tgamma}) is the immediate consequence of the proof of Proposition  2.1
enclosed in Appendix A as the following connections between derivatives and transformation
operators hold:
\begin{eqnarray}
& \partial^{\dagger \;\; \alpha} fg=(\partial^{\dagger \;\; \alpha}f)g
+(\zeta^{- \;\; \alpha}_{\beta}f)\partial^{\dagger \;\; \beta}g & \label{cond2} \\
& (\zeta^{-})^{-1}=\zeta & \label{cond3} \\
& (\partial^{\dagger})^{\dagger \;\; \gamma}=-(-\partial^{\beta}\zeta^{-\;\; \alpha}_{\beta})\zeta^{\gamma}_{\alpha}=\partial^{\gamma}& \label{cond4}
\end{eqnarray}
We notice that $ \zeta^{-} $ is the transformation operator
for the derivative $ \partial^{\dagger}$, the operator $ \zeta $ being its inverse. Therefore all the
calculations from the proof of the Proposition  2.1
can be repeated with suitable replacements. In this way we obtain
the formula for the $ \tilde{\Gamma}$ operator (\ref{tgamma}) which is simply the (\ref{gamma}) operator with new derivatives $ \partial^{\dagger} $
and $ \stackrel{\leftarrow}{\partial}$ and the operator $ \zeta $ acting now as the inverse transformation
operator. 
\\
We modify the operator $ \tilde{\Gamma}$ similarly as in (\ref{hgamma}) and obtain the operator
$ \hat{\tilde{\Gamma}} $:
\begin{equation}
              \hat{\tilde{\Gamma}}_{\mu} (\partial^{\dagger}, \stackrel{\leftarrow}{\partial})=
\stackrel{\leftarrow}{\zeta}^{ j}_{\mu}(\zeta^{\alpha}_{j}\Lambda_{\alpha}) +
\label{thgamma}
\end{equation}
$$
+\sum_{l=1}^{\tilde{N}-1} \sum_{k=0}^{l}
\stackrel{\leftarrow}{\partial}^{ \mu_{1}}...
\stackrel{\leftarrow}{\partial}^{ \mu_{k}}
\stackrel{\leftarrow}{\zeta}^{ j}_{\mu}
\left ( \zeta^{\alpha}_{j}\zeta^{\alpha_{k}}_{\mu_{k}}... \zeta^{\alpha_{1}}_{\mu_{1}} \tilde{\Lambda}_{\alpha_{1}...\alpha_{k} \alpha \mu_{k+1}...\mu_{l}}\right )
\partial^{\dagger \;\; \mu_{k+1}}... \partial^{\dagger \;\; \mu_{l}}
$$
which in turn fufills the equality:
\begin{equation}
\sum_{\mu}  \partial^{\dagger \;\; \mu}\left (f
              \hat{\tilde{\Gamma}}_{\mu} (\partial^{\dagger}, \stackrel{\leftarrow}{\partial})g \right )=
\sum_{\mu} f \left ( (-\stackrel{\leftarrow}{\partial}^{\mu} +\partial^{\dagger \;\; \mu} ) \circ
\tilde{\Gamma}_{\mu} (\partial^{\dagger}, \stackrel{\leftarrow}{\partial})\right )g
\label{thgamma1}
\end{equation}
for a pair of arbitrary functions $ f $ and $ g $.\\
We can use the constructed operators $ \hat{\Gamma}$ and $ \hat{\tilde{\Gamma}} $
to derive the conserved currents by the following proposition:
\begin{four}
Let us assume that the  function $ \Phi $ is an arbitrary solution
of equation (\ref{eqbis}) with coefficients fulfilling (\ref{condconj}) which means:
\begin{equation}
   [\Lambda(\partial) + \tilde{\Lambda}(\partial^{\dagger})] \Phi=0
\end{equation}
and the function $ \Phi' $ solves the conjugated equation
with operators (\ref{eqconj}, \ref{eqconj1}):
\begin{equation}
\Phi' [\Lambda(\stackrel{\leftarrow}{\partial}^{\dagger}) + \tilde{\Lambda}(\stackrel{\leftarrow}{\partial})] =0
\end{equation}
Then:
\begin{equation}
J_{\mu}=\Phi' \hat{\Gamma}_{\mu} \Phi \hspace{2cm} \tilde{J}_{\mu}=\Phi' \hat{\tilde{\Gamma}}_{\mu} \Phi
\label{currents}
\end{equation}
where the operators $\hat{\Gamma}$ and $ \hat{\tilde{\Gamma}} $ are given by
(\ref{hgamma},\ref{thgamma}) is a current that
obeys the conservation law:
\begin{equation}
              \sum_{\mu}\partial^{\mu}J_{\mu}+\sum_{\mu}\partial^{\dagger \;\; \mu}\tilde{J}_{\mu}=0
              \label{cons1}
\end{equation}
\end{four}
Proof:\\
The conservation law (\ref{cons1}) is implied by the properties (\ref{hgamma1}, \ref{thgamma1}) of the applied operators
 $\hat{\Gamma}$ and $ \hat{\tilde{\Gamma}} $:
 \begin{equation}
 \partial^{\mu}\Phi' \hat{\Gamma}_{\mu} \Phi+   \partial^{\dagger \;\; \mu} \Phi' \hat{\tilde{\Gamma}}_{\mu} \Phi=
 \end{equation}
 $$
 \Phi' \left (\sum_{\mu} (-\stackrel{\leftarrow}{\partial}^{\dagger \;\; \mu} +\partial^{\mu} ) \circ
\Gamma_{\mu} (\partial, \stackrel{\leftarrow}{\partial}^{\dagger})+
\sum_{\mu} (-\stackrel{\leftarrow}{\partial}^{\mu} +\partial^{\dagger \;\; \mu} ) \circ
\tilde{\Gamma}_{\mu} (\partial^{\dagger}, \stackrel{\leftarrow}{\partial})\right )
  \Phi= $$
  $$ \Phi' \left (\tilde{\Lambda}(\partial^{\dagger})- \tilde{\Lambda}(\stackrel{\leftarrow}{\partial})
  +\Lambda(\partial)- \Lambda(\stackrel{\leftarrow}{\partial}^{\dagger})
     \right) \Phi=0
$$

\begin{five}
The components of the conserved current for (\ref{eqbis})
fulfilling the condition (\ref{condconj}) can be written also in the form:
\begin{equation}
              J'_{\mu}=\Phi' \hat{\Gamma}_{\mu} \Phi- \zeta^{-\;\; \nu}_{\mu}\left (
              \Phi' \hat{\tilde{\Gamma}}_{\nu} \Phi
              \right )
              \label{current'}
\end{equation}
They obey the conservation law including only the $ \partial$ derivatives:
\begin{equation}
            \sum_{\mu}  \partial^{\mu} J'_{\mu}=0
              \label{cons11}
\end{equation}
provided $ \Phi $ is the solution of (\ref{eqbis}) and the function
$ \Phi' $ solves the conjugate equation with operators (\ref{eqconj},\ref{eqconj1}).
\end{five}
Let us notice that the presented construction holds when we work within the framework
of classical commutative differential calculus (for which the conjugation means
$ \partial^{\dagger}=-\partial $ and the transformation operator is the identity)
 as well as for mixed models where part of the coordinates is continuous
 and the other part is discrete with suitable difference derivatives in the form (\ref{discrete})
 or (\ref{discrete1}). \\
 In the sequel we shall use the construction (\ref{hgamma}, \ref{thgamma}, \ref{currents}) in order to derive the conserved currents
  and in consequence the conserved charges for mixed discrete nonlinear Toda model and for double discrete
  Toda model as well. \\
  As we know from the classical field theory we can use the conserved currents in construction of the
  conserved charges. This is also the case for our discrete and mixed discrete models
  with derivatives defined by (\ref{discrete}) or (\ref{discrete1}). We shall integrate
  the time component of the conserved current
  over subspace excluding the time-coordinate. We use in derivation the conserved currents
  given by
  Proposition 3.2 or accordingly by the Corollary 3.3 if the conjugated derivative with respect
  to time-coordinate appears in the model. Integrating we must remember to take the corresponding
  integral namely for continuous coordinates we understand $ \int dx_{i} $ as the Lebesque integral
  while for the discrete derivative (\ref{discrete}) the definite integral is given by:
  \begin{equation}
  \int dx_{i}:=\sum_{k=-\infty}^{+\infty} (\zeta_{i})^{k}
  \label{integral}
  \end{equation}
  If the equation includes the discrete derivative of the type (\ref{discrete1})
  we use the definite integral in the form:
\begin{equation}
\int dx_{i}:=\sum_{k=-\infty}^{+\infty} (x^{i}_{k+1}-x^{i}_{k})(\zeta_{i})^{k}
\label{integral1}
\end{equation}
In the above formulas for discrete integrals the $ \zeta_{i} $ is the shift operator
in the direction $ i $ along the lattice. \\
Let us denote  the integral over subspace with excluded time-coordinate as $ \int_{sub} $.
Then the conserved charges for discrete and mixed discrete models with variable coefficients
can be derived using the following  proposition:
\begin{six}
Let us assume that in the model described by the equation (\ref{eqbis})
the conjugated discrete derivative with respect to the time-coordinate
does not appear. Then
the charge
\begin{equation}
Q=\int_{sub} J_{t}
\label{charge}
\end{equation}
where $ J_{t} $ is the time component of the conserved current described
in the Proposition 3.2 is conserved:
\begin{equation}
\partial^{t} Q=0
\end{equation}
\end{six}
Proof: \\
The conservation of the charge (\ref{charge}) is implied by the conservation law from
Proposition 3.2 namely:
\begin{equation}
\partial^{t} Q=\int_{sub} \partial^{t} J_{t} =\int_{sub}-\sum_{k \neq t} (\partial^{k} J_{k}+\partial^{\dagger \;\; k}\tilde{J}_{k})=
\end{equation}
$$ =\rm boundary \;\;\; terms=0 $$
provided the currents vanish at the infinity in the corresponding space-like dimensions.
We have used here also the property of the discrete and mixed discrete differential calculus:
\begin{equation}
\partial^{t} \int_{sub}=\int_{sub} \partial^{t}
\label{sub}
\end{equation}
Let us point out that the equality (\ref{sub}) does not apply in general to noncommutative spaces
(the superspace being an exception) and this fact is the major difficulty
in construction of the conserved charges for an arbitrary model on noncommutative space \cite{c3,c4}.
\\
If in the equation (\ref{eqbis}) both initial and conjugate time-derivatives appear
we derive the conserved charge using the Corollary 3.3 namely the following statement is then valid
(with proof being the copy of the above calculations for Proposition 3.4):
\begin{seven}
Let us assume that in the model described by the equation (\ref{eqbis})
both discrete derivatives $ \partial^{t} $ and $ \partial^{\dagger \;\; t} $
appear. Then
the charge
\begin{equation}
Q=\int_{sub} J'_{t}
\label{charge'}
\end{equation}
where $ J'_{t} $ is the time component of the conserved current described
in the Corollary 3.3 is conserved:
\begin{equation}
\partial^{t} Q=0
\end{equation}
\end{seven}
\section{Applications}
\subsection{The second order equation on quantum plane with variable coefficients}
Let us start with the simple example of the second order equation on quantum plane
with coefficients depending on coordinates $ x $ and $ y $. \\
The commutation relations for derivatives and coordinates look as follows \cite{n}:
\begin{equation}
yx=qxy \hspace{2cm} \partial^{y}\partial^{x}=q^{-1} \partial^{x}\partial^{y}
\label{quant}
\end{equation}
The Leibnitz rule (\ref{leib}) in differential calculus is defined by the following $ R $-matrix:
\begin{equation}
R=\left [ \begin{tabular}{cccc}
$ q^{2}$ & 0 & 0 & 0 \\
0 & $q$ &$ q^{2}-1$ & 0 \\
0 & 0 & $q$ & 0  \\
0 & 0 & 0 &$ q^{2}$
\end{tabular} \right ] \hspace{2cm} R'=q^{-2} R
\end{equation}
Form the above matrix we can deduce the explicit action of the transformation operator
$ \zeta $ on the monomials of the first order (\ref{for2}). It yields the following formulas for the inverse transformation
$ \zeta^{-} $:
\begin{eqnarray}
\zeta^{-\;\; x}_{x} x=q^{-2}x & \zeta^{-\;\; x}_{x} y=q^{-1}y \hspace{0,5cm}
\zeta^{-\;\; x}_{y} x=(q^{-2}-1)y & \zeta^{-\;\; x}_{y} y=0 \\
\zeta^{-\;\; y}_{y} y=q^{-2}y & \zeta^{-\;\; y}_{y} x =q^{-1}x \hspace{0,5cm}
\zeta^{-\;\; y}_{x}y=0 &
\zeta^{-\;\; y}_{x} x=0
\end{eqnarray}
The conjugated derivatives are defined using the inverse transformation operator:
\begin{equation}
\partial^{\dagger \;\; x}=-\partial^{x}\zeta^{- \;\; x}_{x}-\partial^{y}\zeta^{- \;\; x}_{y}
\hspace{2cm}
\partial^{\dagger \;\; y}=-\partial^{x}\zeta^{- \;\; y}_{x}-\partial^{y}\zeta^{- \;\; y}_{y}
\end{equation}
Now let us discuss an arbitrary equation of the second order with the coefficients depending
on $ x $ and $ y $ and check which of them fufill the condition (\ref{cond}):
\begin{equation}
\Lambda(\partial^{x},\partial^{y})=\Lambda_{xx}\partial^{x}\partial^{x}
+\Lambda_{yx}\partial^{y}\partial^{x}+\Lambda_{xy}\partial^{x}\partial^{y}
+\Lambda_{yy}\partial^{y}\partial^{y}
\end{equation}
The condition (\ref{cond}) for our simple case reads as follows:
\begin{equation}
\partial^{\dagger \;\; x}\Lambda_{xx} + \partial^{\dagger \;\; y}\Lambda_{yx}=0
\hspace{2cm}
\partial^{\dagger \;\; y}\Lambda_{yy}+\partial^{\dagger \;\; x}\Lambda_{xy}=0
\end{equation}
Let us assume $ \Lambda_{xy}=\Lambda_{yx}=0 $. Then we can show that $ \Lambda_{xx}=\phi(y) $
and $ \Lambda_{yy}=\psi(x) $ obey the conditions (\ref{cond}), namely:
\begin{equation}
\partial^{\dagger \;\; x}\Lambda_{xx}=
(-\partial^{x}\zeta^{- \;\; x}_{x}-\partial^{y}\zeta^{- \;\; x}_{y})\phi(y)=
-\partial^{x}\zeta^{- \;\; x}_{x}\phi(y)=-\partial^{x}\phi(q^{-1}y)=0
\end{equation}
\begin{equation}
\partial^{\dagger \;\; y}\Lambda_{yy}=
(-\partial^{x}\zeta^{- \;\; y}_{x}-\partial^{y}\zeta^{- \;\; y}_{y})\psi(x)=
-\partial^{y}\zeta^{- \;\; y}_{y}\psi(x)=-\partial^{y}\psi(q^{-1}x)=0
\end{equation}
Now we can apply our construction to obtain the $ \Gamma $ operator:
\begin{equation}
\Gamma_{x}={\stackrel{\leftarrow}{\partial}}^{\dagger \;\; x}\phi(q^{-2}y)
+\phi(q^{-1}y)\partial^{x}
\hspace{1cm}
\Gamma_{y}={\stackrel{\leftarrow}{\partial}}^{\dagger \;\; y}\psi(q^{-2}x)
+\psi(q^{-1}x)\partial^{y}
\end{equation}
It is easy to check that this operator fulfills the operator equation (\ref{cond1}):
$$
(-{\stackrel{\leftarrow}{\partial}}^{\dagger \;\; x}+\partial^{x})\circ \Gamma_{x}
+(-{\stackrel{\leftarrow}{\partial}}^{\dagger \;\; y}+\partial^{y})\circ \Gamma_{y} $$
\begin{equation}
= -\left ({\stackrel{\leftarrow}{\partial}}^{\dagger \;\; x}{\stackrel{\leftarrow}{\partial}}^{\dagger \;\; x}
\phi(q^{-2}y)+
{\stackrel{\leftarrow}{\partial}}^{\dagger \;\; y}{\stackrel{\leftarrow}{\partial}}^{\dagger \;\; y}
\psi(q^{-2}x) \right )+
 \phi(y)\partial^{x}\partial^{x} +\psi(x) \partial^{y}\partial^{y}
\end{equation}
where the operator $ \Lambda(\stackrel{\leftarrow}{\partial} ) $ gives the conjugated equation:
\begin{equation}
\Lambda(\stackrel{\leftarrow}{\partial} )= {\stackrel{\leftarrow}{\partial}}^{\dagger \;\; x}{\stackrel{\leftarrow}{\partial}}^{\dagger \;\; x}
\phi(q^{-2}y)+
{\stackrel{\leftarrow}{\partial}}^{\dagger \;\; y}{\stackrel{\leftarrow}{\partial}}^{\dagger \;\; y}
\psi(q^{-2}x)
\end{equation}
The modified $ \hat{\Gamma} $ operator is given by the formula:
\begin{eqnarray}
& \hat{\Gamma}_{x} ={\stackrel{\leftarrow}{\partial}}^{\dagger \;\; x}
{\stackrel{\leftarrow}{\zeta}}^{-\;\; x}_{x}\phi(q^{-2}y)
+{\stackrel{\leftarrow}{\zeta}}^{-\;\; x}_{x}\phi(q^{-1}y)\partial^{x}
 & \\
& \hat{\Gamma}_{y} ={\stackrel{\leftarrow}{\partial}}^{\dagger \;\; x}
{\stackrel{\leftarrow}{\zeta}}^{-\;\; x}_{y}\phi(q^{-2}y)
+{\stackrel{\leftarrow}{\zeta}}^{-\;\; x}_{y}\phi(q^{-1}y)\partial^{x}
& \nonumber \\
&
+{\stackrel{\leftarrow}{\partial}}^{\dagger \;\; y}
{\stackrel{\leftarrow}{\zeta}}^{-\;\; y}_{y}\psi(q^{-2}x)
+{\stackrel{\leftarrow}{\zeta}}^{-\;\; y}_{y}\psi(q^{-1}x)\partial^{y}
 &
\end{eqnarray}
Having the explicit form of the $ \Gamma $ and $ \hat{\Gamma} $ operators we can construct the conserved currents:
\begin{equation}
J_{x}=\Phi' \hat{\Gamma}_{x} \Phi \hspace{2cm}  J_{y}=\Phi' \hat{\Gamma}_{y} \Phi
\end{equation}
where the functions $ \Phi' $ and $ \Phi $ solve the respective equations:
\begin{eqnarray}
& \Phi' \left ( {\stackrel{\leftarrow}{\partial}}^{\dagger \;\; x}{\stackrel{\leftarrow}{\partial}}^{\dagger \;\; x}
\phi(q^{-2}y)+
{\stackrel{\leftarrow}{\partial}}^{\dagger \;\; y}{\stackrel{\leftarrow}{\partial}}^{\dagger \;\; y}
\psi(q^{-2}x) \right )=0 & \label{qeq} \\
& \left (\phi(y)\partial^{x}\partial^{x} +\psi(x) \partial^{y}\partial^{y}
\right )\Phi=0 &           \label{qeq1}
\end{eqnarray}
According to the Proposition 2.2 the above current obeys the conservation law:
\begin{equation}
\partial^{x}J_{x}+\partial^{y}J_{y}=0
\end{equation}
provided functions $ \Phi' $ and $ \Phi $ are the solutions of the corresponding equations (\ref{qeq}, \ref{qeq1}).
\subsection{Nonlinear equation of motion for chiral and antichiral supermultiplets}
The supersymmetric models in the superfield formulation yield interesting examples of the equations of motion
with coefficients depending on variables \cite{buc,sri}. Let us recall that in such a framework
the fields depend on superspace coordinates including the space-time and spinor
variables while in construction of the action the covariant derivatives are used which explicitly depend
on spinor coordinates.\\
We shall study in this section the D=4 N=1 chiral and antichiral
superfields obeying the nonlinear equation of motion resulting from the supersymmetric action:
\begin{equation}
I=\int d^{4}xd^{2}\theta d^{2}\bar{\theta} \bar{\Phi} \Phi + \int d^{4}xd^{2}\theta \left ( \frac{m}{2}\Phi^{2}
+\frac{g}{3}\Phi^{3}\right ) +\int d^{4}xd^{2}\bar{\theta}\left ( \frac{m}{2}{\bar{\Phi}}^{2}
+\frac{g}{3}{\bar{\Phi}}^{3}\right )
\label{action}
\end{equation}
The integration over spinor variables $ \theta $ and $\bar{\theta}$ can be expressed in terms of the covariant derivatives:
\begin{eqnarray}
& \int d^{2} \theta =-\frac{1}{4} D^{2} \hspace{1cm}   \int d^{2} \bar{\theta} =-\frac{1}{4} {\bar{D}}^{2} & \\
& \int d^{2}\theta d^{2} \bar{\theta} =\frac{1}{16} D^{2}{\bar{D}}^{2}= \frac{1}{16}{\bar{D}}^{2}{D}^{2}
=\frac{1}{16} D^{\alpha} {\bar{D}}^{2} D_{\alpha} =\frac{1}{16} {\bar{D}}_{
\dot \alpha}D^{2}{\bar{D}}^{\dot \alpha} &
\end{eqnarray}
The covariant derivatives  used in supersymmetric models are built
from the basic derivatives with respect to the space-time and spinor coordinates:
\begin{eqnarray}
              & D_{\alpha} =\partial_{\alpha}+i \sigma^{\mu}_{\alpha \dot \beta}{\bar{\theta}}^{\dot \beta} \partial_{\mu}
              \hspace{1cm}   D^{\alpha}=\epsilon^{\alpha \beta} D_{\beta} & \label{cov1}\\
              &  \bar{D}_{\dot \alpha} =-\partial_{\dot \alpha}-i \theta^{\beta}\sigma^{\mu}_{ \beta \dot \alpha} \partial_{\mu}
              \hspace{1cm}   \bar{D}^{\dot \alpha}=\epsilon^{\dot \alpha \dot \beta} \bar{D}_{\dot \beta} & \label{cov2}
\end{eqnarray}
Due to the fact that chiral $ \Phi$ and antichiral $ \bar{\Phi} $ superfields
fulfill the following condition:
\begin{equation}
\bar{D}_{\dot \alpha} \Phi=0 \hspace{2cm}  D_{\alpha} \bar{\Phi}=0
\end{equation}
we obtain from the action (\ref{action}) the equations of motion in the form:
\begin{equation}
   \left (\begin{tabular}{cc}
   $ \frac{1}{4} D^{2}$  &$  m+g \bar{\Phi}$ \\
   $ m+g \Phi $& $ \frac{1}{4} \bar{D}^{2}$
   \end{tabular} \right )  \left ( \begin{tabular}{c} $ \Phi$ \\$ \bar{\Phi} $
   \end{tabular} \right )=0
   \label{eqsusy}
\end{equation}
We have written the equations for the superfields $ \Phi $ and $ \bar{\Phi} $ in the matrix form. In
this example we deal with the coefficients of the equation depending explicitly on $ \theta $ and $ \bar{\theta}$
variables  due to the form of the covariant derivatives (\ref{cov1}, \ref{cov2}):
\begin{eqnarray}
& D^{2} =D^{\alpha} D_{\alpha}=-\partial^{\alpha}\partial_{\alpha} +2i {\bar{\theta}}^{\dot \beta}
\sigma^{\mu}_{\alpha \dot \beta} \partial_{\mu}\partial^{\alpha} -{\bar{\theta}}^{2}\square
  & \\
  & \bar{D}^{2} =\bar{D}_{\dot \alpha} \bar{D}^{\dot \alpha}=-\bar{\partial}_{\dot \alpha}\bar{\partial}^{\dot \alpha} -2i \theta^{ \alpha}
\sigma^{\mu}_{\alpha \dot \beta} \partial_{\mu}\bar{\partial}^{\dot \beta} +\theta^{2}\square
  &
\end{eqnarray}
The second important feature of the chiral-antichiral superfield  equation of motion is its nonlinearity
for $ g \neq 0 $. However the nonlinear term does not depend on the derivatives so we can apply our method
of derivation the conservation law provided the condition (\ref{cond}) for the kinetic part of the operator is fulfilled. Let
us extract the $ \zeta $ and $ \zeta^{-} $ operators from the Leibnitz rule for basic derivatives:
\begin{eqnarray}
& \partial^{\mu} (fg)=(\partial^{\mu}f)g +f\partial^{\mu} g & \\
& \partial^{\alpha} (fg)=(\partial^{\alpha}f)g +(\zeta^{\alpha}_{\beta}f)\partial^{\beta} g & \\
& \bar{\partial}^{\dot \alpha} (fg)=(\bar{\partial}^{\dot \alpha}f)g +( \zeta^{\dot \alpha}_{\dot \beta} f)\bar{\partial}^{\dot \beta} g &
\end{eqnarray}
It is clear that $ \zeta^{\mu}_{\nu}=\delta^{\mu}_{\nu} $ and the following
components of the transformation operator vanish:
\begin{equation}
 \zeta^{\mu}_{\alpha}=\zeta^{\alpha}_{\mu}=0 \hspace{1cm}
 \zeta^{\mu}_{\dot \alpha}=\zeta^{\dot \alpha}_{\mu}=0 \hspace{1cm}
 \zeta^{\dot \alpha}_{\alpha}=\zeta^{\alpha}_{\dot \alpha}=0
\end{equation}
The remaining components of the $ \zeta $ operator are defined by its action on the monomials of the first order in superspace
variables:
\begin{equation}
\zeta^{\alpha}_{\beta} x_{\nu}=\delta^{\alpha}_{\beta} x_{\nu} \hspace{1cm}
\zeta^{\alpha}_{\beta} \theta_{\gamma}=-\delta^{\alpha}_{\beta} \theta_{\gamma} \hspace{1cm}
\zeta^{\alpha}_{\beta} \bar{\theta}_{\dot \gamma}=- \delta^{\alpha}_{\beta} \bar{\theta}_{\dot \gamma}
\end{equation}
The analogous formulas hold also for dotted indices. Hence we get for the inverse transformation operator:
\begin{equation}
\zeta^{- \;\; \alpha}_{\beta} x_{\nu}=\delta^{\alpha}_{\beta} x_{\nu} \hspace{1cm}
\zeta^{- \;\; \alpha}_{\beta} \theta_{\gamma}=-\delta^{\alpha}_{\beta} \theta_{\gamma} \hspace{1cm}
\zeta^{- \;\; \alpha}_{\beta} \bar{\theta}_{\dot \gamma}=- \delta^{\alpha}_{\beta} \bar{\theta}_{\dot \gamma}
\end{equation}
with the corresponding expressions with dotted indices of the same form.\\
Now having the explicitly derived inverse transformation operator we arrive at the conjugated derivatives
given by the formulas:
\begin{equation}
\partial^{\dagger \;\; \mu} =-\partial^{\mu} \hspace{1cm}
\partial^{\dagger \;\; \alpha}=-\partial^{\beta} \zeta^{- \;\; \alpha}_{\beta} \hspace{1cm}
\bar{\partial}^{\dagger \;\; \dot \alpha}=-\bar{\partial}^{\dot \beta} \zeta^{- \;\; \dot \alpha}_{\dot \beta}
\end{equation}
Using the conjugated derivatives we obtain the condition (\ref{cond}) written for our equation for the chiral and antichiral
superfields:
\begin{eqnarray}
& \partial^{\dagger \;\; \alpha}\Lambda_{\alpha \mu}+
\bar{\partial}^{\dagger\;\; \dot \alpha}\Lambda_{\dot \alpha \mu}+
\partial^{\dagger \;\; \nu}\Lambda_{\nu \mu}=0  & \\
& \partial^{\dagger \;\; \alpha} \Lambda_{\alpha  \beta}+\partial^{\dagger \;\; \mu}\Lambda_{\mu \beta}=0  & \\
& \bar{\partial}^{\dagger \;\; \dot \alpha} \Lambda_{\dot \alpha \dot  \beta}+\partial^{\dagger \;\; \mu}\Lambda_{\mu\dot \beta}=0 &
\end{eqnarray}
One can easily check that the above conditions are fulfilled using the explicit form of the
coefficients from equation (\ref{eqsusy}):
\begin{eqnarray}
              & \Lambda_{\alpha \dot \beta}=\Lambda_{\dot \beta \alpha}=0 & \\
              & \Lambda_{\alpha \beta}=-\epsilon_{\alpha \beta} \left (
              \begin{tabular}{cc}
              $\frac{1}{4}$ & 0 \\
              0 & 0\end{tabular} \right )
              & \\
              & \Lambda_{\dot \alpha \dot \beta}=-\epsilon_{\dot \alpha\dot \beta} \left (
              \begin{tabular}{cc}
              $0 $ & 0 \\
              0 &$ \frac{1}{4}$ \end{tabular} \right ) & \\
              & \Lambda_{\mu \alpha}=\Lambda_{\alpha \mu} =\left (
              \begin{tabular}{cc}
              $\frac{i}{4}\bar{\theta}^{\dot \beta}\sigma^{\nu}_{\alpha \dot \beta}$ & 0 \\
              0 & 0\end{tabular} \right )g_{\nu \mu} & \\
              & \Lambda_{\mu \dot \alpha}=\Lambda_{\dot \alpha \mu} =\left (
              \begin{tabular}{cc}
              0 & 0 \\
              0 & $ -\frac{i}{4}\theta^{\beta}\sigma^{\nu}_{\beta \dot \alpha}$
              \end{tabular} \right )g_{\nu \mu} &           \\
              & \Lambda_{\mu \nu} = \left ( \begin{tabular}{cc}
              $ - \frac{1}{4} \bar{\theta}^{2}$ & 0 \\
              0 & $  -\frac{1}{4} \theta^{2}$
              \end{tabular}\right ) g_{\mu \nu}
\end{eqnarray}
The next step is therefore the  application of the general formula (\ref{gamma}) for the $ \Gamma $ operator
to our model:
\begin{eqnarray}
& \Gamma_{\mu} = {\stackrel{\leftarrow}{\partial}}^{\dagger \;\; \nu}\Lambda_{\nu \mu}
+ \Lambda_{\mu \nu}\partial^{\nu} & \\
& \Gamma_{\alpha} = -{\stackrel{\leftarrow}{\partial}}^{\dagger \;\; \mu}\Lambda_{\mu \alpha}
+ {\stackrel{\leftarrow}{\partial}}^{\dagger \;\; \gamma}\Lambda_{\gamma \alpha}
 +\Lambda_{\alpha \gamma}\partial^{\gamma}-\Lambda_{\alpha \mu} \partial^{\mu} & \\
 & \Gamma_{\dot \alpha} = -{\stackrel{\leftarrow}{\partial}}^{\dagger \;\; \mu}\Lambda_{\mu \dot \alpha}
+ {\stackrel{\leftarrow}{\partial}}^{\dagger \;\; \dot \gamma}\Lambda_{\dot \gamma \dot \alpha}
 +\Lambda_{\dot \alpha \dot \gamma}\partial^{\dot \gamma}-\Lambda_{\dot \alpha \mu} \partial^{\mu} &
 \end{eqnarray}
 The operator $ \hat{\Gamma}$ differs from the $ \Gamma $ by insertion of the $ \zeta^{-} $ operator in the middle
 of the monomials of derivatives:
\begin{eqnarray}
& \hat{\Gamma}_{\mu}=\Gamma_{\mu} & \\
& \hat{\Gamma}_{\beta}=
 -{\stackrel{\leftarrow}{\partial}}^{\dagger \; \mu} {\stackrel{\leftarrow}{\zeta}}^{- \; \alpha}_{\beta}\Lambda_{\mu \alpha}
+ {\stackrel{\leftarrow}{\partial}}^{\dagger \; \gamma} {\stackrel{\leftarrow}{\zeta}}^{- \; \alpha}_{\beta}\Lambda_{\gamma \alpha}
 + {\stackrel{\leftarrow}{\zeta}}^{- \; \alpha}_{\beta}\Lambda_{\alpha \gamma}\partial^{\gamma}
 - {\stackrel{\leftarrow}{\zeta}}^{- \; \alpha}_{\beta}\Lambda_{\alpha \mu} \partial^{\mu} & \\
 & \hat{\Gamma}_{\dot \beta} =
-{\stackrel{\leftarrow}{\partial}}^{\dagger \; \mu}{\stackrel{\leftarrow}{\zeta}}^{- \; \dot \alpha}_{\dot \beta}\Lambda_{\mu \dot \alpha}
+ {\stackrel{\leftarrow}{\partial}}^{\dagger \; \dot \gamma}{\stackrel{\leftarrow}{\zeta}}^{- \; \dot \alpha}_{\dot \beta}\Lambda_{\dot \gamma \dot \alpha}
 +{\stackrel{\leftarrow}{\zeta}}^{- \; \dot \alpha}_{\dot \beta}\Lambda_{\dot \alpha \dot \gamma}\partial^{\dot \gamma}
 -{\stackrel{\leftarrow}{\zeta}}^{- \; \dot \alpha}_{\dot \beta}\Lambda_{\dot \alpha \mu} \partial^{\mu} &
\end{eqnarray}
We apply the obtained operator $ \hat{\Gamma} $ to derive the currents:
\begin{eqnarray}
& J_{\mu}= \left ( \bar{\Phi}', \Phi' \right )\hat{\Gamma}_{\mu} \left (
\begin{tabular}{c} $ \Phi$ \\$ \bar{\Phi} $
   \end{tabular} \right ) & \label{csusy1} \\
&   J_{\alpha}= \left ( \bar{\Phi}', \Phi' \right )\hat{\Gamma}_{\alpha} \left (
\begin{tabular}{c} $ \Phi$ \\$ \bar{\Phi} $
   \end{tabular} \right ) & \label{csusy2} \\
& J_{\dot \alpha}= \left ( \bar{\Phi}', \Phi' \right )\hat{\Gamma}_{\dot \alpha} \left (
\begin{tabular}{c} $ \Phi$ \\$ \bar{\Phi} $
   \end{tabular} \right )     &   \label{csusy3}
\end{eqnarray}
which obey the conservation law:
\begin{equation}
\partial^{\mu}J_{\mu}+\partial^{\alpha} J_{\alpha} + \bar{\partial}^{\dot \alpha} J_{\dot \alpha}=0
\end{equation}
provided the fields used in construction fulfill the corresponding equation of motion (\ref{eqsusy})
and the pair of superfields $ \bar{\Phi}' $ and $ \Phi' $ its conjugated version:
\begin{equation}
              \left ( \bar{\Phi}', \Phi'\right ) \left (
              \begin{tabular}{cc}
   $ \frac{1}{4} \stackrel{\leftarrow}{D^{\dagger}}^{2} $  &$  m+g \bar{\Phi}$ \\
   $ m+g \Phi $& $ \frac{1}{4} \stackrel{\leftarrow}{\bar{D}^{\dagger}}^{2} $
   \end{tabular} \right )=0
\end{equation}
Let us notice that this equation is linear with respect to superfields $ \bar{\Phi}' $and
$ \Phi' $ and its solution depends on the potential given by the chiral and antichiral superfields $ \Phi $ and $ \bar{\Phi} $.\\
Due to the properties of covariant derivatives built form the selfconjugated operators
we conclude that the  conjugated set of equations for superfields $(\bar{\Phi}', \Phi' ) $
has at least following solutions:
\begin{itemize}
\item{for the case $ g \neq 0 $ we can take $\bar{\Phi}'=\Phi$ and $\Phi'=\bar{\Phi} $
where $ \Phi $ and $ \bar{\Phi} $ are solutions for (\ref{eqsusy}),}
\item{when $ g=0 $ the superfield $ \bar{\Phi}'$ is an arbitrary solution of chiral and $ \Phi' $
of antichiral part of the set of equations (\ref{eqsusy}).}
\end{itemize}
The above solutions give in turn the explicit form of the conserved currents (\ref{csusy1}-\ref{csusy3})
where we replace the multiplet $(\bar{\Phi}', \Phi' ) $
with $ (\Phi,\bar{\Phi} ) $ for $ g \neq 0 $
and with $ (\delta \Phi,\delta \bar{\Phi} )$ for $ g=0 $. Let us notice that for
linear model $ g=0 $ we obtain the full set of conserved currents connected
with symmetry operators of chiral-antichiral supermultiplet equation:
\begin{eqnarray}
& J_{\mu}= \left (\delta \Phi,\delta \bar{\Phi} \right )\hat{\Gamma}_{\mu} \left (
\begin{tabular}{c} $ \Phi$ \\$ \bar{\Phi} $
   \end{tabular} \right ) &  \\
&   J_{\alpha}= \left (\delta\Phi,\delta\bar{\Phi} \right )\hat{\Gamma}_{\alpha} \left (
\begin{tabular}{c} $ \Phi$ \\$ \bar{\Phi} $
   \end{tabular} \right ) &  \\
& J_{\dot \alpha}= \left (\delta \Phi,\delta \bar{\Phi}  \right )\hat{\Gamma}_{\dot \alpha} \left (
\begin{tabular}{c} $ \Phi$ \\$ \bar{\Phi} $
   \end{tabular} \right )     &
\end{eqnarray}
where the symmetries $ \delta $ include supersymmetric transformations:
\begin{equation}
Q_{\alpha}=-i\partial_{\alpha}-\sigma^{l}_{\alpha \dot \beta }\bar{\theta}^{\dot \beta}\partial_{l}
\hspace{1cm}
\bar{Q}^{\dot \alpha}=-i\bar{\partial}^{\dot \alpha}+\sigma^{l \;\; \dot \alpha}_{\alpha}\theta^{ \alpha}\partial_{l}
\end{equation}
and the operators from Poincar\`{e} algebra: momenta, angular momentum and boosts
for four-dimensional Minkowski space. \\
We can use the obtained conserved current (\ref{csusy1} - \ref{csusy3}) to construct the integral of motion by integrating
the time-component of the current (\ref{csusy1}) over the subspace:
\begin{equation}
Q=\int d^{3} x d^{2} \theta d^{2} \bar{\theta} J_{0}
\end{equation}
as the following equality holds:
\begin{equation}
\partial^{0} Q =\int d^{3}x d^{2} \theta d^{2}\bar{\theta} \;\; \partial^{0}J_{0}=
\end{equation}
$$
\int d^{3}x d^{2} \theta d^{2}\bar{\theta}\left ( -\partial^{k}J_{k}-\partial^{\alpha}J_{\alpha}-\partial^{\dot \alpha}J_{\dot \alpha}\right )=\rm
boundary \;\;\; terms=0
$$
Let us point out that the developed method allows immediate construction
of conserved charges. The obtained charges are supermultiplets
built from component charges which are also conserved separately. The application of our method
to the equations obeying the condition (\ref{cond})
is an alternative to the procedure used in supersymmetric models
(see for example \cite{su1} for chiral superfields and \cite {ehm}
for supersymmetric principal chiral model) where the conservation laws include
covariant spinor derivatives $ D $ and $\bar{D} $
instead of basic derivatives $ \partial^{\mu}$, $ \partial^{\alpha} $
and $ \bar{\partial}^{\dot \alpha} $ which we have used. The consequence
of the fact that the time-derivative does not appear explicitly
in the conservation law is an additional procedure required to derive
the component conservation laws and conserved charges.
\subsection{Nonlinear integrable models and their consistency condition}
\subsubsection{Principal chiral model equation}
We shall discuss the application of the construction given by Propositions  2.1 and 2.2
 to the classical problem of nonlocal conserved
currents and charges for the principal chiral model. The model is built
using the commutative differential calculus, nevertheless its linearized version
yields an interesting example of the linear equation with coefficients fulfilling the
condition (\ref{cond}). Dimakis and M\"{u}ller-Hoissen \cite{mh1,mh2} treated this case within their framework
of the gauged bi-covariant noncommutative differential calculus and obtained  the linearized
version of the equations in the form:
\begin{eqnarray}
& \partial^{t} \chi =\lambda (\partial^{x}+g^{-1}(\partial^{x}g)) \chi & \label{set1}\\
& \partial^{x} \chi =\lambda (\partial^{t}+g^{-1}(\partial^{t}g)) \chi & \label{set2}
\end{eqnarray}
where $ \chi $ is an $ N \times N $ matrix of smooth functions depending on $ t $ and $ x $ while
the field $ g $ denotes the principal chiral field and the initial nonlinear equation
of the principal chiral model  looks as follows:
\begin{equation}
\partial^{x}(g^{-1}\partial^{x}g)=\partial^{t}(g^{-1}\partial^{t}g)
\label{pchm}
\end{equation}
 In the paper \cite{mh2}
the integrability condition for the set (\ref{set1},\ref{set2}) was used and it appeared that it is first
order in derivatives therefore yields the conservation law with currents linear with respect
to the field $ \chi $ and depending on the $ \lambda $ parameter. \\
We propose in addition to consider the consistency  condition derived from the set (\ref{set1},\ref{set2})
which has the following form:
\begin{equation}
              \left [\partial^{x}\partial^{x} -\partial^{t} \partial^{t}
              +g^{-1}(\partial^{x}g)\partial^{x}-
              g^{-1}(\partial^{t}g)\partial^{t}\right ]\chi=0
              \label{ceq3}
\end{equation}
As we see the operator of this equation is free of the parameter
$ \lambda$ and the condition (\ref{cond}) becomes identical with the principal chiral model equation (\ref{pchm}):
\begin{equation}
              \partial^{\dagger \;\; x}\Lambda_{x}
              +\partial^{\dagger \;\; t}\Lambda_{t}=
              \partial^{x}g^{-1}(\partial^{x}g)-
              \partial^{t}g^{-1}(\partial^{t}g)=0
\end{equation}
the remaining coefficients $ \Lambda $ of the operator $ \Lambda(\partial) $ being constant. \\
The conjugated equation for (\ref{ceq3}) together with connected conjugated set for
(\ref{set1}, \ref{set2}) look as follows:
\begin{eqnarray}
       & \chi'\left [\stackrel{\leftarrow}{\partial}^{x}\stackrel{\leftarrow}{\partial}^{x}
        -\stackrel{\leftarrow}{\partial}^{t}\stackrel{\leftarrow}{ \partial}^{t}
              -\stackrel{\leftarrow}{\partial}^{x}g^{-1}(\partial^{x}g)+
              \stackrel{\leftarrow}{\partial}^{t}g^{-1}(\partial^{t}g)\right ]=0 & \label{ceqconj3} \\
& -\partial^{t} \chi' =\lambda \left [-\partial^{x}\chi'+\chi'g^{-1}(\partial^{x}g) \right] & \label{set1conj}\\
& -\partial^{x} \chi' =\lambda \left [-\partial^{t}\chi'+\chi'g^{-1}(\partial^{t}g) \right ]& \label{set2conj}
\end{eqnarray}
We work in this example with classical commutative basic derivatives so the transformation operator is simply the identity operator and the operators $ \Gamma $ and
$ \hat{\Gamma} $ coincide:
\begin{eqnarray}
              & \hat{\Gamma}_{x}=\Gamma_{x}=g^{-1}(\partial^{x}g)-
              \stackrel{\leftarrow}{\partial}^{x}+\partial^{x} & \\
              &  \hat{\Gamma}_{t}=\Gamma_{t}=-g^{-1}(\partial^{t}g)+
              \stackrel{\leftarrow}{\partial}^{t}-\partial^{t} & \label{gammat}
\end{eqnarray}
Our consistency condition allows the construction of the conserved current according to the Proposition 2.2.
The conserved current is given by the derived operator $ \Gamma$ and the solution $ \chi $ of the equation (\ref{ceq3})
while $ \chi' $ denotes the solution of the conjugated equation (\ref{ceqconj3}-\ref{set2conj}):
\begin{equation}
              J_{x} =\chi'\Gamma_{x} \delta \chi
              \hspace{2cm}   J_{t} =\chi'\Gamma_{t} \delta \chi
\end{equation}
where we have denoted as $ \delta $ the symmetry operator of the equation (\ref{ceq3}).
The operator $ \delta $ can be derived using the extended vector fields
method described in \cite{ol}. We have checked that the following operators transform solution
$ \chi $ of the consistency condition (\ref{ceq3}) into solution $ \delta \chi$:
\begin{equation}
              \delta^{0}=C \frac{\partial}{\partial \chi}
              \hspace{0,7cm}
              \delta^{1}=\chi \frac{\partial}{\partial \chi}
              \hspace{0,7cm}
              \delta^{2}=g^{-1} \frac{\partial}{\partial \chi}
              \hspace{0,7cm}
              \delta^{3}=(\partial^{x})^{-1} (g^{-1}\partial^{t}g)\frac{\partial}{\partial \chi}
\end{equation}
where $ C $ is a constant matrix $ N \times N $. \\
The above symmetry operators of the equation (\ref{ceq3}) imply the following currents:
\begin{equation}
J^{0}_{\mu}=\chi'\Gamma_{\mu} C \hspace{0,5cm}
J^{1}_{\mu}=\chi'\Gamma_{\mu} \chi \hspace{0,5cm}
J^{2}_{\mu}=\chi'\Gamma_{\mu}g^{-1} \hspace{0,5cm}
J^{3}_{\mu}=\chi'\Gamma_{\mu} (\partial^{x})^{-1} (g^{-1}\partial^{t}g)
\label{currents3}
\end{equation}
which are conserved:
\begin{equation}
              \partial^{x}J^{i}_{x}+\partial^{t}J^{i}_{t}=0
\end{equation}
thereby yielding the conserved charges after the integration of the time-components of the curents (\ref{currents3})
with respect to the space variable $ x$:
\begin{eqnarray}
& Q^{0}=\int dx \;\; \chi'\Gamma_{t} C & \\
& Q^{1}=\int dx \;\;  \chi'\Gamma_{t} \chi & \\
& Q^{2}=\int dx \;\;  \chi'\Gamma_{t}g^{-1} & \\
& Q^{3}=\int dx \;\;  \chi'\Gamma_{t} (\partial^{x})^{-1} (g^{-1}\partial^{t}g) &
\end{eqnarray}
where the operator $ \Gamma_{t} $ is given by formula (\ref{gammat}). \\
Now following \cite{mh2} we can expand the fields $ \chi $ and $ \chi' $ in terms of powers of the parameter
$ \lambda$ and obtain the infinite set of conserved charges connected with the nonlinear equation
(\ref{pchm}) of the principal chiral model:
\begin{equation}
              \chi=\sum_{m=0}^{\infty} \lambda^{m}\chi^{(m)}
              \hspace{1cm}
              \chi'=\sum_{m=0}^{\infty} \lambda^{m}\chi'^{(m)}
              \label{exp}
\end{equation}
Using equations (\ref{set1}, \ref{set2})
and (\ref{set1conj},\ref{set2conj}) we arrive at the following set of equations for coefficients
$ \chi^{(m)} $ and $ \chi'^{(m)}$ (we assume after \cite{mh2} that $ \chi^{(0)}$,
$ \chi^{(1)}$,  $ \chi'^{(0)}$ and
$ \chi'^{(1)}$ are the unital $ N \times N $ matrices):
\begin{eqnarray}
\partial^{t} \chi^{(m)}=[\partial^{x} +g^{-1}(\partial^{x}g)]\chi^{(m-1)} & & m\ge 2 \\
\partial^{x} \chi^{(m)}=[\partial^{t} +g^{-1}(\partial^{t}g)]\chi^{(m-1)} & & m\ge 2 \\
-\partial^{t} \chi'^{(m)}=-\partial^{x}\chi'^{(m-1)} +\chi'^{(m-1)}g^{-1}(\partial^{x}g) & & m\ge 2 \\
-\partial^{x} \chi'^{(m)}=-\partial^{t}\chi'^{(m-1)} +\chi'^{(m-1)}g^{-1}(\partial^{t}g) & & m\ge 2
\end{eqnarray}
which give after solution:
\begin{eqnarray}
              & \chi^{(2)}=(\partial^{x})^{-1} g^{-1}(\partial^{t}g)=(\partial^{t})^{-1} g^{-1}(\partial^{x}g)=-\chi'^{(2)} & \\
              & \chi^{(m)} =\left [(\partial^{x})^{-1}(\partial^{t}+g^{-1}(\partial^{t}g)
              \right]^{m-2}(\partial^{x})^{-1}g^{-1}(\partial^{t}g) & \label{sol1} \\
              & \chi'^{(m)}= (-1)^{m-2}\chi'^{(2)}\left [\left (-{\stackrel{\leftarrow}{\partial}}^{t}
              +g^{-1}(\partial^{t}g)\right )(\stackrel{\leftarrow}{\partial^{x}})^{-1}\right ]^{m-2} &                                     \label{sol2}
\end{eqnarray}
Inserting the solution (\ref{sol1},\ref{sol2}) into the conserved currents (\ref{currents3}) we obtain after expansion with respect to the parameter $ \lambda$
the infinite tower of conserved charges for the field $ g $ of the principal chiral model:
\begin{eqnarray}
            & Q^{0 \;\; (m)}=\int dx \;\; \chi'^{(m)}\Gamma_{t} C & \\
            & Q^{1 \;\; (m)}=\int dx \;\;  \sum_{k=0}^{m} \chi'^{(k)}\Gamma_{t} \chi^{(m-k)} & \\
            & Q^{2 \;\; (m)}=\int dx  \;\; \chi'^{(m)}\Gamma_{t}g^{-1} & \\
            & Q^{3 \;\; (m)}=\int dx  \;\; \chi'^{(m)}\Gamma_{t} (\partial^{x})^{-1} g^{-1}(\partial^{t}g)&
\end{eqnarray}
\subsubsection{Generalization of the principal chiral model equation}
We can easily extend the results of the previous section to the case of the principal
chiral model with the fields depending smoothly on variables $ t $, $ x $ and $ y $.
Let us recall the linearized version for $ N \times N $ matrices $ \chi $ with  entries
depending on the mentioned variables \cite{mh2}:
\begin{eqnarray}
 & \partial^{t} \chi =\lambda (\partial^{x}+g^{-1}(\partial^{x}g)) \chi & \label{set1'}\\
& \partial^{y} \chi =\lambda (\partial^{t}+g^{-1}(\partial^{t}g)) \chi &  \label{set2'}
\end{eqnarray}
with the conjugated set looking as follows:
\begin{eqnarray}
& \partial^{t} \chi' =\lambda \left [\partial^{x}\chi'-\chi'g^{-1}(\partial^{x}g) \right] & \label{cset1'}\\
& \partial^{y} \chi' =\lambda \left [\partial^{t}\chi'-\chi'g^{-1}(\partial^{t}g) \right ]& \label{cset2'}
\end{eqnarray}
The integrability condition  of the set (\ref{set1'},\ref{set2'}) yields the currents
discussed in \cite{mh2}. Following the previous example we propose to investigate
the consistency condition for (\ref{set1'},\ref{set2'}) which becomes:
\begin{equation}
      \left [\partial^{x}\partial^{y} -\partial^{t} \partial^{t}
              +g^{-1}(\partial^{x}g)\partial^{y}-
              g^{-1}(\partial^{t}g)\partial^{t}\right ]\chi=0
              \label{ceq4}
\end{equation}
with its conjugation in the sense (\ref{eqconj}) of the form:
 \begin{equation}
      \left [\partial^{x}\partial^{y} -\partial^{t} \partial^{t}\right ]\chi'
              +\chi'\left[-\stackrel{\leftarrow}{\partial}^{y} g^{-1}(\partial^{x}g)+
              \stackrel{\leftarrow}{\partial}^{t}g^{-1}(\partial^{t}g)\right ]=0
              \label{ceqconj4}
\end{equation}
The principal chiral model equation coincides with  the condition (\ref{cond}) for variable coefficients:
\begin{equation}
 \partial^{\dagger \;\; y}\Lambda_{y}
              +\partial^{\dagger \;\; t}\Lambda_{t}=
              \partial^{y}g^{-1}(\partial^{x}g)-
              \partial^{t}g^{-1}(\partial^{t}g)=0
              \label{pchm'}
\end{equation}
We can write down the components of the $ \Gamma $ operator taking into account that
due for the fact of the basic derivatives being commutative,
the transformation operator is the identity so the
operators $ \Gamma $ and $ \hat{\Gamma} $ coincide:
\begin{eqnarray}
               & \hat{\Gamma}_{x}=\Gamma_{x}= \frac{1}{2}\left (- \stackrel{\leftarrow}{\partial}^{y}+\partial^{y} \right ) & \\
               & \hat{\Gamma}_{y}=\Gamma_{y}=g^{-1}(\partial^{x}g)
              +\frac{1}{2} \left (-\stackrel{\leftarrow}{\partial}^{x}+\partial^{x} \right )& \\
              &  \hat{\Gamma}_{t}=\Gamma_{t}=-g^{-1}(\partial^{t}g)+
              \stackrel{\leftarrow}{\partial}^{t}-\partial^{t} &
\end{eqnarray}
The current built using the above operators and the solutions $ \chi $ and $ \chi' $ of the
respective equations (\ref{set1'} - \ref{cset2'}) looks as follows:
\begin{equation}
              J^{\delta}_{x} =\chi'\Gamma_{x} \delta \chi
              \hspace{1cm} J^{\delta}_{y}=\chi' \Gamma_{y} \delta \chi
              \hspace{1cm}   J^{\delta}_{t} =\chi'\Gamma_{t} \delta \chi
              \label{currents'}
\end{equation}
with $ \delta $ being the symmetry operator from the set:
\begin{equation}
   \delta^{0}=C \frac{\partial}{\partial \chi}
   \hspace{0,5cm}
  \delta^{1}=\chi \frac{\partial}{\partial \chi}
              \hspace{0,5cm}
              \delta^{2}=g^{-1} \frac{\partial}{\partial \chi}
              \hspace{0,5cm}
              \delta^{3}=(\partial^{y})^{-1} (g^{-1}\partial^{t}g)\frac{\partial}{\partial \chi}
\end{equation}
where $ C $ is a constant $ N \times N $ matrix. \\
According to the Proposition 2.2 the currents (\ref{currents'}) obey the conservation law:
\begin{equation}
              \partial^{t}J^{i}_{t}+\partial^{x}J^{i}_{x}+\partial^{y}J^{i}_{y}=0
\end{equation}
Thus the currents (\ref{currents'}) produce the infinite tower of conserved charges similarly to the previous
case. After expansion of the fields $ \chi $ and $ \chi'$ in terms of powers of the parameter
 $ \lambda $ (\ref{exp}) we obtain:
 \begin{eqnarray}
    & Q^{0 \;\; (m)}=\int dxdy \chi'^{(m)} \Gamma_{t} C & \\
    & Q^{1 \;\; (m)}=\int dxdy  \;\; \sum_{k=0}^{m} \chi'^{(k)}\Gamma_{t} \chi^{(m-k)} & \\
            & Q^{2 \;\; (m)}=\int dxdy \;\;   \chi'^{(m)}\Gamma_{t}g^{-1} & \\
            & Q^{3 \;\; (m)}=\int dxdy  \;\; \chi'^{(m)}\Gamma_{t} (\partial^{y})^{-1} g^{-1}(\partial^{t}g)&
\end{eqnarray}
where the explicit expressions for the component fields given by the set of equations (\ref{set1'} - \ref{cset2'})
look as follows ( with $ \chi^{(0)}$,
$ \chi^{(1)}$,  $ \chi'^{(0)}$ and
$ \chi'^{(1)}$ being the unital $ N \times N $ matrices as before):
\begin{eqnarray}
& \chi^{(2)}=-\chi'^{(2)}=(\partial^{y})^{-1}g^{-1}(\partial^{t}g) & \\
& \chi^{(m)}=\left [(\partial^{y})^{-1}(\partial^{t}+g^{-1}(\partial^{t}g))
\right]^{m-2}(\partial^{y})^{-1}g^{-1}(\partial^{t}g) & m\ge 2 \\
& \chi'^{(m)}=(-1)^{m-2}\chi'^{(2)}\left [\left (-{\stackrel{\leftarrow}{\partial}}^{t}
              +g^{-1}(\partial^{t}g)\right )(\stackrel{\leftarrow}{\partial^{y}})^{-1}\right ]^{m-2} & m\ge 2
\end{eqnarray}
\subsubsection{Nonlinear Toda lattice equation}
The interesting case of application of the Proposition 3.2 is the linear set of equations
which originates from nonlinear Toda lattice equation. The set of auxiliary equations derived in the paper \cite{mh2}
from gauged bicovariant differential calculus
reads as follows:
\begin{eqnarray}
& \dot \chi_{k}=\lambda (e^{q_{k-1}-q_{k}}\chi_{k-1}-\chi_{k} ) \zeta & \label{tset1}\\
& \chi_{k+1}-\chi_{k}=-\lambda (\dot \chi_{k}+\dot q_{k}\chi_{k} )\zeta & \label{tset2}
\end{eqnarray}
where we have used our notation for the transformation operator
which for the mixed model like the one considered in this section is simply
shifting along the lattice $ \zeta f_{k}=f_{k-1} $ with respect to the space-like dimension.
 Again we shall discuss the consistency
condition which for the set of equations (\ref{tset1},\ref{tset2}) becomes an  equation of second order
including classical derivative with respect to the time-variable and the discrete one
with respect to the space-lattice variable:
\begin{equation}
              \ddot \chi_{k}+\dot q_{k}\dot \chi_{k} -e^{q_{k-1}-q_{k}}(\chi_{k-1}-\chi_{k})
              -(\chi_{k+1}-\chi_{k})=0
\end{equation}
Let us notice that this equation is free of the parameter $ \lambda $ and the operator $ \zeta $. \\
We can rewrite this equation using the notation:
\begin{eqnarray}
\chi_{k-1}-\chi_{k}=(\zeta -1)\chi_{k}=\partial\chi_{k} & & \chi_{k+1}-\chi_{k}=(\zeta^{-}-1)\chi_{k}=\partial^{\dagger} \chi_{k} \\
q_{k-1}-q_{k}=(\zeta -1) q_{k}=\partial q_{k} & & \dot f=\partial^{t} f
\end{eqnarray}
and it has the following form:
\begin{equation}
\left [\partial^{t}\partial^{t} +(\partial^{t}q)\partial^{t}
-e^{(\partial q)}\partial -\partial^{\dagger}\right]\chi=0
\label{ceq5}
\end{equation}
In this form we see clearly that the equation involves both $ \partial $ and $ \partial^{\dagger} $ derivatives
so we should follow the construction for discrete models given in Proposition 3.1 and 3.2. \\
Let us observe that the discussed equation fufills the restriction (\ref{cond}) and it coincides
in this case with the nonlinear Toda lattice equation:
\begin{equation}
\partial^{\dagger \;\; t} \Lambda_{t}
+\partial^{\dagger} \Lambda_{x}=-\ddot q_{k}-e^{q_{k}-q_{k+1}}+e^{q_{k-1}-q_{k}}=0
\end{equation}
We construct the $ \Gamma $ operator according to (\ref{gamma}, \ref{hgamma}, \ref{tgamma}, \ref{thgamma}) taking into account that the transformation operator acts as follows:
\begin{eqnarray}
\zeta_{t} f_{k}(t)=f_{k}(t) &
\zeta f_{k}(t)=f_{k-1}(t) &
\zeta^{-} f_{k}(t)=f_{k+1}(t) \\
             & \hat{\Gamma}_{t}=\Gamma_{t}=-{\stackrel{\leftarrow}{\partial}}^{t}
             +\partial^{t}+(\partial^{t} q) & \\
             & \hat{\Gamma}_{x}=-{\stackrel{\leftarrow}{\zeta}}^{-} e^{-(\partial^{\dagger} q)} & \\
             & \hat{\tilde{\Gamma}}_{x}={\stackrel{\leftarrow}{\zeta}}
\end{eqnarray}
The components of the $ \hat{\Gamma} $ operator yield the corresponding components
of the current:
\begin{equation}
J^{\delta}_{t}=\delta\chi'\hat{\Gamma}_{t} \chi \hspace{1cm}
J^{\delta}_{x}=\delta\chi'\hat{\Gamma}_{x}  \chi \hspace{1cm}
\tilde{J}^{\delta}_{x}=\delta\chi'\hat{\tilde{\Gamma}}_{x}  \chi
\end{equation}
which obey the conservation law including the conjugated derivative:
\begin{equation}
              \partial^{t}J^{\delta}_{t}+\partial J^{\delta}_{x}+\partial^{\dagger} \tilde{J}^{\delta}_{x}=0
\end{equation}
provided the field $ \chi $ fufills the consistency condition (\ref{ceq5}), the field $ \chi' $ its conjugation:
\begin{equation}
              \chi' \left [\stackrel{\leftarrow}{\partial}^{t}\stackrel{\leftarrow}{\partial}^{t}
              -\stackrel{\leftarrow}{\partial}^{t}(\partial^{t}q)-\stackrel{\leftarrow}{\partial}^{\dagger}
              e^{-(\partial^{\dagger})q}-\stackrel{\leftarrow}{\partial}
              \right ]=0
              \label{ceqconj5}
\end{equation}
and the operator $ \delta $ is now the symmetry operator of the conjugated equation
(\ref{ceqconj5}).\\
We proceed taking integrals with respect to the discrete spatial variable (we have denoted it as $ x$)
and obtain the  charges:
\begin{equation}
Q^{\delta}=\int dx \;\; J_{t}=\int dx \;\;  \delta\chi' \hat{\Gamma}_{t}  \chi
\label{acharge}
\end{equation}
which are also conserved:
\begin{equation}
              \partial^{t} Q=0
\end{equation}
In the above formula we understand the discrete definite integral  $ \int dx $ as
given by (\ref{integral}). \\
Analogously to the examples for principal chiral model we can check that the following operators transform
solution $ \chi' $ of the equation (\ref{ceqconj5}) into the solution $ \delta \chi' $:
\begin{equation}
              \delta^{0}=c \frac{\partial}{\partial \chi'}
              \hspace{1cm}
              \delta^{1}=e^{q} \frac{\partial}{\partial \chi'}
              \hspace{1cm}
              \delta^{2}=[-\partial^{-1} (\partial^{t} q)+t]\frac{\partial}{\partial \chi'}
\end{equation}
where the indefinite discrete integral $\partial^{-1} $ is given by:
\begin{equation}
              \partial^{-1} =\frac{1}{\zeta-1}=-\sum_{k=0}^{\infty}\zeta^{k}
              \label{indef}
\end{equation}
and $ c $ is a constant. \\
The derived symmetries of the equation (\ref{ceq5}) lead to the following expressions for charges:
\begin{eqnarray}
              & Q^{0}=\int dx \;\; c \Gamma_{t} \chi & \\
              & Q^{1}=\int dx \;\; e^{q} \Gamma_{t}\chi & \\
              & Q^{2}=\int dx \;\; [-\partial^{-1} (\partial^{t} q)+t] \Gamma_{t} \chi &
\end{eqnarray}
After expanding the solutions $ \chi $  with respect to powers of the operators $ \lambda \zeta $
and $ \chi' $ in terms of $ \lambda \stackrel{\leftarrow}{\zeta}^{-} $ we shall be able to rewrite the expressions
(\ref{acharge}) as infinite set of conserved charges:
\begin{eqnarray}
& Q^{0(m)}=\int dx \;\;  c\Gamma_{t} \chi^{(m)} & \\
&  Q^{1(m)}=\int dx  \;\; e^{q}\Gamma_{t}\chi^{(m)} & \\
&  Q^{2(m)}=\int dx  \;\; [-\partial^{-1} (\partial^{t} q)+t] \Gamma_{t}\chi^{(m)}
\end{eqnarray}
where the components $ \chi^{(m)} $ read as follows:
\begin{equation}
\chi^{(m)}=(-1)^{m-1} \left [ (\partial^{\dagger})^{-1}(\partial^{t}+(\partial^{t}q))
\right ]^{m-2}(\partial^{\dagger})^{-1} (\partial^{t} q) \hspace{1cm} m \ge 2
\end{equation}
\subsubsection{Nonlinear Toda lattice equation - generalization}
Our next example is the generalization of nonlinear Toda model
sketched in \cite{mh2}. The functions depend now smoothly on one time variable $ t $, one spacelike
variable $ y $ and one discrete space-like coordinate $ x $. The bicovariant differential calculus
is in this case defined as follows \cite{mh2}:
\begin{eqnarray}
              \delta f=[S^{-1},f] \tau-f' \xi & & df=\dot f \tau +[S,f]\xi \\
              &A=X \tau +(Y-1) S\xi &
\end{eqnarray}
where $ X $ and $ Y $ are matrices with entries being smooth functions of $ t $ and $ y $ and discrete in the coordinate $ x $. The
condition $ \delta A =0 $ leads to the following generalizd Toda model nonlinear equations:
\begin{equation}
X'_{k}=Y_{k}-Y_{k-1} \hspace{1cm} \dot Y_{k} =Y_{k}X_{k+1}-X_{k}Y_{k}
\label{Toda}
\end{equation}
The linear version for auxiliary field $ \chi $ can be deduced following \cite{mh2} from the set
of equation given by:
\begin{equation}
              \delta \chi =\lambda (d+A) \chi
\end{equation}
and looks as follows:
\begin{eqnarray}
              & (S^{-1}-1)\chi=\lambda [\dot \chi+X \chi] S & \label{tset1'}\\
              & -\chi'=\lambda[(S-1)\chi +(Y-1)( S\chi)]S &   \label{tset2'}
\end{eqnarray}
We can rewrite the above set of equations using our notations:
\begin{eqnarray}
           S=\zeta & & s^{-1}=\zeta^{-} \\
           \zeta \chi_{k}(t,y)=\chi_{k+1} (t,y)& &   \zeta^{-} \chi_{k}(t,y)=\chi_{k-1}(t,y) \\
              (\zeta-1)\chi=\partial \chi & &  (\zeta^{-}-1)\chi=\partial^{\dagger} \chi\\
              \chi'=\partial^{y} \chi & & \dot \chi=\partial^{t} \chi
\end{eqnarray}
and they are given by the formulas:
\begin{eqnarray}
              & \partial^{\dagger} \chi=\lambda[\partial^{t}+X] \chi \zeta & \label{tset1n} \\
              & -\partial^{y} \chi=\lambda [\partial \chi +(Y-1)(\zeta\chi) ]  \zeta &  \label{tset2n}
\end{eqnarray}
The consistency equation for the above set of equations reads as:
\begin{equation}
              [\partial^{y}\partial^{t}+X\partial^{y} -Y\partial -\partial^{\dagger}]\chi=0
              \label{ceq6}
\end{equation}
Let us check the condition (\ref{cond}) for the variable coefficients of the operator of the equation:
\begin{equation}
              \partial^{\dagger \;\; y}\Lambda_{y}+\partial^{\dagger}\Lambda_{x}=
              -\partial^{y}X-Y+\zeta^{-} Y=0
\end{equation}
As before  our equation fulfills the restriction (\ref{cond}) due to the
fact that $ X $ and $ Y $ obey the generalization of the nonlinear Toda model (\ref{Toda}).
The condition (\ref{condconj}) also holds as the coefficient for the derivative $ \partial^{\dagger} $
is constant.\\
Similarly to the previous examples we can apply the Propositions 3.1 and 3.2 for the discrete models
depending on initial and conjugate discrete derivatives. In this way
we obtain the $ \hat{\Gamma} $ operator in the form:
\begin{eqnarray}
              \hat{\Gamma}_{t}=\Gamma_{t}=\frac{1}{2} (-{\stackrel{\leftarrow}{\partial}}^{y}+\partial^{y} ) & &
              \hat{\Gamma}_{y}=\Gamma_{y}=\frac{1}{2} (-{\stackrel{\leftarrow}{\partial}}^{t}+\partial^{t} ) +X \\
              \hat{\Gamma}_{x}=-{\stackrel{\leftarrow}{\zeta}}^{-} (\zeta^{-}Y) & &
              \hat{\tilde{\Gamma}}_{x}=-\stackrel{\leftarrow}{\zeta}
\end{eqnarray}
We use the above components of the $ \hat{\Gamma} $ and $ \hat{\tilde{\Gamma}}$ operators to construct
the respective components of the currents:
\begin{eqnarray}
              J_{t}^{\delta}=\delta \chi'\hat{\Gamma}_{t}  \chi & & J_{y}^{\delta}=\delta \chi'\hat{\Gamma}_{y}  \chi \\
              J_{x}^{\delta}=\delta \chi'\hat{\Gamma}_{x}  \chi & & \tilde{J}_{x}^{\delta}=\delta \chi'\hat{\tilde{\Gamma}}_{x}  \chi
\end{eqnarray}
where  we have denoted $ \delta $ as the symmetry operator of the conjugated equation (\ref{conjeq6})
and $ \chi $ is the solution of (\ref{tset1n} - \ref{ceq6}) while $ \chi' $ solves its conjugation:
\begin{equation}
              [-\partial^{y}\partial^{t} +\partial]\chi'
              +(\partial^{y} \chi')X +(\partial^{\dagger} \chi') (\zeta^{-} Y)=0
              \label{conjeq6}
\end{equation}
As the currents fulfill the conservation law:
\begin{equation}
              \partial^{t}J_{t}^{\delta}+\partial^{y}J_{y}^{\delta}
              +\partial J_{x}^{\delta}+\partial^{\dagger}\tilde{J}_{x}^{\delta}=0
\end{equation}
they yield the conserved charges (with the integral $ \int dx $ in the sense of (\ref{integral})):
\begin{equation}
              Q^{\delta}=\int dy  dx \;\;\; J_{t}^{\delta} =\int dy  dx \;\;\; \delta \chi'\hat{\Gamma}_{t}  \chi
\end{equation}
Following the procedure applied earlier we expand the solutions $ \chi $ and $ \chi'$
in terms of powers of the operator $ \lambda \zeta $ and $ \lambda {\stackrel{\leftarrow}{\zeta}}^{-} $
respectively:
\begin{equation}
              \chi=\sum_{m=0}^{\infty} (\lambda \zeta )^{m}\chi^{(m)}
              \hspace{1cm} \chi'=\sum_{m=0}^{\infty} (\lambda \stackrel{\leftarrow}{\zeta}^{-} )^{m}\chi'^{(m)}
              \label{exp'}
\end{equation}
After assuming the first components in the form of the unital $ N \times N $ matrix:
\begin{equation}
              \chi^{(0)} =\chi'^{(0)}=\chi^{(1)}=\chi'^{(1)}= {\bf 1}_{N \times N }
\end{equation}
we obtain the following equations for the subsequent components:
\begin{eqnarray}
              \partial^{\dagger} \chi^{(m)}=[\partial^{t}+X] \chi^{(m-1)} & & m \ge 2\\
              -\partial^{y} \chi^{(m)}= [\partial +(Y-1)\zeta ] \chi^{(m-1)} & & m \ge 2
\end{eqnarray}
Inserting the expansion (\ref{exp'}) into the expressions for currents and charges
we arrive at the infinite towers of the conserved currents and charges:
\begin{eqnarray}
              & J^{\delta \;\; (m)}_{\mu}= \delta\chi' \hat{\Gamma}_{\mu} \chi^{(m)} & \\
              & Q^{\delta \;\; (m)} = \int dxdy \;\;\;  \delta\chi'  \hat{\Gamma}_{t} \chi^{(m)} &
\end{eqnarray}
with the following explicit expressions for the components of the fields $ \chi $:
\begin{eqnarray}
              & \chi^{(2)}=\chi'^{(2)} =(\partial^{\dagger})^{-1} X & \\
              & \chi^{(m)}=[ (\partial^{\dagger})^{-1}(\partial^{t}+X)]^{m-2} (\partial^{\dagger})^{-1} X & m \ge 2
\end{eqnarray}
and the initial solutions for the conjugated equation (\ref{conjeq6})
related to the symmetry operators in the form:
\begin{equation}
      \delta^{0}\chi'=C \hspace{1cm} \delta^{1}\chi'=\partial^{-1} X-y
\end{equation}
with $ C $ being a constant $ N \times N $ matrix.
\subsubsection{Double discrete nonlinear Toda lattice equation}
Let us end the review of the integrable models with fully discrete Toda lattice equation.
The set of auxiliary linear equations given in \cite{mh1} has the following form:
\begin{eqnarray}
& \chi_{k}(n)-\chi_{k}(n-1)=-\frac{\gamma}{c} [g^{-1}_{k}(n)g_{k+1}(n)\chi_{k+1}(n)-\chi_{k}(n) ] & \label{s1}\\
& \chi_{k}(n)-\chi_{k-1}(n)=-\gamma c [g^{-1}_{k}(n)g_{k}(n+1)\chi_{k}(n+1)-\chi_{k}(n) ] &         \label{s2}
\end{eqnarray}
We rewrite the above equations using our notation:
\begin{eqnarray}
f_{k}(n)-f_{k-1}(n)=-\partial^{x} f_{k}(n) & & \zeta^{-}_{x} f_{k}(n)=f_{k+1}(n) \\
 f_{k}(n)-f_{k}(n-1)=-\partial^{t} f_{k}(n) & & \zeta^{-}_{t}f_{k}(n)=f_{k}(n+1)
\end{eqnarray}
Then the set of equations (\ref{s1},\ref{s2}) looks as follows:
\begin{eqnarray}
& -\partial^{t}\chi =-\frac{\gamma}{c}[g^{-1}(\zeta^{-}_{x}g)(\zeta^{-}_{x}\chi)-\chi] &\\
& -\partial^{x}\chi =-\gamma c[g^{-1}(\zeta^{-}_{t}g)(\zeta^{-}_{t}\chi)-\chi] &
\end{eqnarray}
The consistency equation for the above set reads as:
\begin{equation}
-g^{-1}(\zeta^{-}_{x}g)\partial^{\dagger \;\; x} \chi -\partial^{x} \chi
+c^{2}g^{-1}(\zeta^{-}_{t}g)\partial^{\dagger \;\; t} \chi +c^{2}\partial^{t} \chi=0
\label{conjdis}
\end{equation}
The condition (\ref{condconj}) for our example is fulfilled:
\begin{eqnarray}
& \partial^{\dagger \;\; x}\Lambda_{x}+\partial^{\dagger \;\; t}\Lambda_{t}=0 & \\
& \partial^{x}\tilde{\Lambda}_{x}+\partial^{t}\tilde{\Lambda}_{t}=
-\partial^{x} g^{-1}(\zeta^{-}_{x}g)+c^{2} \partial^{t}g^{-1}(\zeta^{-}_{t}g) =0
\end{eqnarray}
provided the double discrete nonlinear Toda lattice equation  is valid for the field $ g $ :
\begin{equation}
  \partial^{x} g^{-1}(\zeta^{-}_{x}g)=c^{2} \partial^{t}g^{-1}(\zeta^{-}_{t}g)
\end{equation}
We construct the conserved currents following Propositions 3.1 and 3.2. As both initial and
conjugated derivatives appear in the equation we use the $ \Gamma $ and $ \tilde{\Gamma} $
operators:
\begin{eqnarray}
              \Gamma_{x}=-1 & & \tilde{\Gamma}_{x}=-(\zeta_{x}g^{-1})g \\
              \Gamma_{t}=c^{2} & & \tilde{\Gamma}_{t}=c^{2}(\zeta_{t}g^{-1})g
\end{eqnarray}
and obtain the $ \hat{\Gamma} $ and $ \hat{\tilde{\Gamma}} $ operators with the following components:
\begin{eqnarray}
              \hat{\Gamma}_{x}=-\stackrel{\leftarrow}{\zeta}^{-}_{x} & & \hat{\tilde{\Gamma}}_{x}=-\stackrel{\leftarrow}{\zeta}_{x}(\zeta_{x}g^{-1})g \\
              \hat{\Gamma}_{x}=c^{2}\stackrel{\leftarrow}{\zeta}^{-}_{t} & & \hat{\tilde{\Gamma}}_{t}=c^{2}\stackrel{\leftarrow}{\zeta}_{t}(\zeta_{t}g^{-1})g
\end{eqnarray}
The above operators give in turn the currents:
\begin{equation}
J^{\delta}_{\mu} = \delta\chi' \hat{\Gamma}_{\mu} \chi \hspace{2cm}
\tilde{J}^{\delta}_{\mu} =\delta \chi' \hat{\tilde{\Gamma}}_{\mu} \chi
\label{currdis}
\end{equation}
which are conserved according to the Proposition 3.2:
\begin{equation}
\partial^{t} J^{\delta}_{t}+\partial^{x} J^{\delta}_{x}
+\partial^{\dagger \;\; t} \tilde{J}^{\delta}_{t}
+\partial^{\dagger \;\; x} \tilde{J}^{\delta}_{x}=0
\end{equation}
provided the function  $ \chi $ solves the equation (\ref{conjdis}),
 $ \chi' $ the conjugated
equation:
\begin{equation}
\chi'\left [ \stackrel{\leftarrow}{\partial}^{x} (\zeta_{x}g^{-1})g+
 \stackrel{\leftarrow}{\partial}^{\dagger \;\; x}
 -c^{2}\stackrel{\leftarrow}{\partial}^{t} (\zeta_{t}g^{-1})g
 -c^{2}\stackrel{\leftarrow}{\partial}^{\dagger \;\; t} \right ]=0
\end{equation}
and $ \delta $ is the symmetry operator for the conjugated equation. \\
Following the Corollary 3.3 we can reformulate the currents (\ref{currdis}) in order to obtain the conservation
law including only the $ \partial $ derivatives:
\begin{equation}
J'^{\delta}_{x}=J^{\delta}_{x}-\zeta^{-}_{x}(\tilde{J}^{\delta}_{x}) \hspace{1cm}
J'^{\delta}_{t}=J^{\delta}_{t}-\zeta^{-}_{t}(\tilde{J}^{\delta}_{t})
\end{equation}
The current $ J' $ obeys the conservation law:
\begin{equation}
              \partial^{t} J'^{\delta}_{t}+\partial^{x} J'^{\delta}_{x}=0
\end{equation}
Now we apply the (\ref{current'}) version of the conserved current so as to obtain the conserved charges:
\begin{equation}
  Q^{\delta}=\int dx J'^{\delta}_{t}= \int dx \left ( \delta\chi' \hat{\Gamma}_{t} \chi -\zeta^{-}_{t}(
  \delta\chi' \hat{\tilde{\Gamma}}_{t} \chi) \right )
\end{equation}
where the integral $ \int dx $ is defined as in (\ref{integral}).
\section{Final remarks}
We have discussed the method of derivation of the conservation laws for a
class of equations with variable coefficients. It can be applied to  models
built using supersymmetric, discrete, mixed discrete and noncommutative
(quantum Minkowski or braided) differential calculus. \\
The conserved charges were constructed explicitly for considered supersymmetric and discrete
equations. The problem of derivation of such charges for noncommutative models
is open. The integration over noncommutative spaces
namely braided linear space (including q-Minkowski)
is well developed \cite{z,q,int,c3}
but subintegrals and commutation rules for subintegrals and derivatives
need further study. In particular the generalized version of the  property (\ref{sub})
$$ \partial^{t} \int_{sub}=\int_{sub} \partial^{t} $$
must be derived. As we have shown for models on quantum planes
\cite{c3} it is crucial in the construction of the conserved charges.
\section{Appendix A}
Proof of the Proposition 2.1: \\
Let us recall the notation for monomials of derivatives:
\begin{equation}
[\rho_{1}...\rho_{k}]:=\partial^{\rho_{1}}...\partial^{\rho_{k}}
\hspace{2cm} \overline{[\nu_{1}...\nu_{m}]}:={\stackrel{\leftarrow}{\partial}}^{\dagger \;\; \nu_{1}}...
{\stackrel{\leftarrow}{\partial}}^{\dagger \;\; \nu_{m}}
\end{equation}
 Due to the modification of the Leibnitz rule
Leibnitz rule  we are to consider the solution of the operator equation
for the $ \Gamma $ operator
 in the form of the polynomial of order $ N-1 $:
\begin{equation}
\Gamma_{\mu} (\partial, \stackrel{\leftarrow}{\partial}^{\dagger})=
a^{0}_{\mu} +  \sum_{l=1}^{N-1} \sum_{k=0}^{l}
\overline{[\mu_{1},...,\mu_{k}]}a^{k}_{\mu \mu_{1}... \mu_{l}}
[\mu_{k+1},...,\mu_{l}]
\end{equation}
where the coefficients $ a^{k} $ depend on the coordinates
$ \vec{x} $. \\
The condition (\ref{cond1}) from the section 2 applied to the above polynomial yields
the equations for coefficients $ a^{k}_{\mu \mu_{1}...\mu_{l}} $:
\begin{equation}
\sum_{\mu} (-\stackrel{\leftarrow}{\partial}^{\dagger \;\; \mu} +\partial^{\mu} ) \circ
\Gamma_{\mu} (\partial, \stackrel{\leftarrow}{\partial}^{\dagger})=
\end{equation}
$$ - \sum_{l=1}^{N-1} \sum_{k=0}^{l} \sum_{\mu}
\overline{[\mu_{1},...,\mu_{k},\mu]}a^{k}_{\mu \mu_{1}... \mu_{l}}
[\mu_{k+1},...,\mu_{l}]  $$
$$ +  \sum_{l=1}^{N-1} \sum_{k=0}^{l}
\overline{[\mu_{1},...,\mu_{k}]}\sum_{\mu}(\zeta^{\mu}_{\nu}a^{k}_{\mu \mu_{1}... \mu_{l}})
[\nu,\mu_{k+1},...,\mu_{l}] $$
$$  + \sum_{l=1}^{N-1} \sum_{k=0}^{l}
\overline{[\mu_{1},...,\mu_{k}]}\sum_{\mu} (\partial^{\mu}a^{k}_{\mu \mu_{1}... \mu_{l}})
[\mu_{k+1},...,\mu_{l}] $$
$$ -\sum_{\mu}\overline{[\mu]}a^{0}_{\mu}
+ \sum_{\mu} (\zeta^{\mu}_{\nu} a^{0}_{\mu} )[\nu]
+\sum_{\mu} (\partial^{\mu} a^{0}_{\mu}) = $$
$$ =\Lambda(\partial)- \Lambda(\stackrel{\leftarrow}{\partial}^{\dagger}) $$
The procedure is analogous to the one used in the proof of the Proposition 4.1
from \cite{c2} which describes the derivation of conservation law
for the equation with constant coefficients or fulfilling the strong condition (see (23) and (24) from \cite{c2}).
Now we have decided to change the form of the conjugated equation. This results
in the weaker restrictions for coefficients of the equation (\ref{eq}) and changes the set
of equations for functions
$ a^{k}_{\mu \mu_{1}...\mu_{l}}$ defining the operator $\Gamma $ for the following one:
\begin{eqnarray}
& \partial^{\mu}a^{0}_{\mu}=0 & \label{eq0} \\
& a^{0}_{\mu}= \zeta^{- \;\; \alpha}_{\mu}\Lambda_{\alpha}  & \label{eq1'} \\
& \partial^{\alpha}a^{0}_{\alpha \mu}+\zeta^{\nu}_{\mu} a^{0}_{\nu}=\Lambda_{\mu}  & \label{eq2}  \\
& \zeta^{\alpha}_{\mu}a^{0}_{\alpha \mu_{1}... \mu_{l}}
+ \partial^{\alpha} a^{0}_{\alpha \mu \mu_{1} ...\mu_{l}}=
\Lambda_{ \mu \mu_{1}...\mu_{l}} &    \label{eq3} \\
 & -a^{k}_{\mu \mu_{1}...\mu_{l}}
+ \zeta^{\alpha}_{\mu_{k+1}} a^{k+1}_{\alpha \mu_{1} ... \mu_{k} \mu \mu_{k+2}... \mu_{l}}
+ \partial^{\alpha} a^{k+1}_{\alpha \mu_{1}... \mu_{k} \mu \mu_{k+1}... \mu_{l}}=
0 &   \label{eq4} \\
& a^{l}_{\mu \mu_{1}... \mu_{l}}+ \partial^{\alpha} a^{l}_{\alpha \mu_{1}...\mu_{l}\mu}=
\zeta^{- \;\; \alpha}_{\mu} \zeta^{- \;\; \alpha_{l}}_{\mu_{l}} ...\zeta^{- \;\; \alpha_{1}}_{\mu_{1}}
\Lambda_{\alpha_{1}...\alpha_{l} \alpha} & \label{eq5}
\end{eqnarray}
with $ l=1,...,N-1 \hspace{1cm}  k=0,...,l-1 $. \\
We see that the equations (\ref{eq0}, \ref{eq1'}, \ref{eq5}) yielded
by the coefficients of the conjugated equation differ from the case studied earlier
in \cite{c1,c2}.
We begin to solve this set of equations by deriving the coefficients
 $ a^{0}_{\mu \mu_{1}...\mu_{l}} $ from equations (\ref{eq1'},\ref{eq2},\ref{eq3}). Namely
for $ l=N-1 $ we have:
\begin{equation}
\zeta^{\alpha}_{\mu} a^{0}_{\alpha \mu_{1}...\mu_{N-1}}=\Lambda_{\mu \mu_{1}...\mu_{N-1}}
\end{equation}
Applying the inverse operator $ \zeta^{-} $ we obtain:
\begin{equation}
a^{0}_{\mu \mu_{1}...\mu_{N-1}}=\zeta^{- \;\; \alpha}_{\mu}\Lambda_{\alpha \mu_{1}...\mu_{N-1}}
\end{equation}
We insert this solution into (\ref{eq3}) for $ l=N-2 $ and solve the next equation :
\begin{equation}
\zeta^{\alpha}_{\mu} a^{0}_{\alpha \mu_{1}...\mu_{N-2}}
-\partial^{\dagger \;\; \alpha}\Lambda_{\alpha \mu \mu_{1}...\mu_{N-2}}=
\Lambda_{\mu \mu_{1}...\mu_{N-2}}
\end{equation}
By assumption (\ref{cond}) after using $ \zeta^{-} $ operator and (\ref{inverse}) we derive
$ a^{0}_{\mu \mu_{1}...\mu_{N-2}} $ as:
\begin{equation}
a^{0}_{\mu \mu_{1}...\mu_{N-2}}=\zeta^{- \;\; \alpha}_{\mu}\Lambda_{\alpha \mu_{1}...\mu_{N-2}}
\end{equation}
Passing to the next equation from the subset (\ref{eq3}) and solving them in
the similar way we obtain the unique solution for coefficients
$ a^{0} $:
\begin{equation}
a^{0}_{ \mu_{1}... \mu_{l}}=\zeta^{- \;\; \alpha}_{\mu_{1}}\Lambda_{\alpha \mu_{2}... \mu_{l}} \hspace{1cm} l=1,...,N
\end{equation}
This  solution for initial coefficients
allows us to evaluate the remaining ones using (\ref{eq4},\ref{eq5}), namely
we obtain the $ a^{1} $ coefficients after writing the subset (\ref{eq4})
for $ k=0 $ and solving it the way we solved the subset (\ref{eq1'},\ref{eq2},\ref{eq3}) for $ a^{0} $.
The result is unique and looks as follows:
\begin{equation}
 a^{1}_{\mu \mu_{1} \mu_{2} ... \mu_{l}}=\zeta^{- \;\; \alpha}_{\mu}\zeta^{- \;\; \alpha_{1}}_{\mu_{1}}
 \Lambda_{\alpha_{1} \alpha \mu_{2}...\mu_{l}}
\hspace{2cm} l=1,...,N-1
 \end{equation}
The same method applied to subsets of (\ref{eq4},\ref{eq5}) for $ k=1,...,N-2 $
produces the unique solution of the set of equations for coefficients in the form:
\begin{equation}
a^{k}_{\mu \mu_{1}...\mu_{l}}=\zeta^{- \;\; \alpha}_{\mu}\zeta^{- \;\; \alpha_{k}}_{\mu_{k}}...
\zeta^{- \;\; \alpha_{1}}_{\mu_{1}}\Lambda_{\alpha_{1}...\alpha_{k} \alpha \mu_{k+1}...\mu_{l}}
\hspace{2cm} l=1,...,N-1
\end{equation}
The derivation of the explicit formulas for unique solution
of the coefficients of the operator $ \Gamma_{\mu} $ concludes the proof of
Proposition 2.1. \\
Let us notice once more that in the derivation of coefficients
for the $ \Gamma_{\mu} $ operator the crucial factors were
the properties of coefficients of the equation (\ref{eq} - \ref{cond})
which enabled us to solve the equation (\ref{cond1}) in the explicit form.

\end{document}